# CHEMSMART: Chemistry Simulation and Modeling Automation Toolkit for High-Efficiency Computational Chemistry Workflows

Xinglong Zhang,[1,*] Huiwen Tan,[1] Jingyi Liu,[1] Zihan Li,[1] Lewen Wang,[2] Benjamin W. J. Chen[3]

[1]*Department of Chemistry, The Chinese University of Hong Kong, Shatin, New Territories, Hong Kong, China*

[2]*Department of Chemistry, Hong Kong Baptist University, Kowloon, Hong Kong, China*

[3]*Institute of High Performance Computing, Agency for Science, Technology and Research (A*STAR), 1 Fusionopolis Way, #16-16, Connexis, Singapore 138632, Singapore*

Email: xinglong.zhang@cuhk.edu.hk

## Graphical TOC

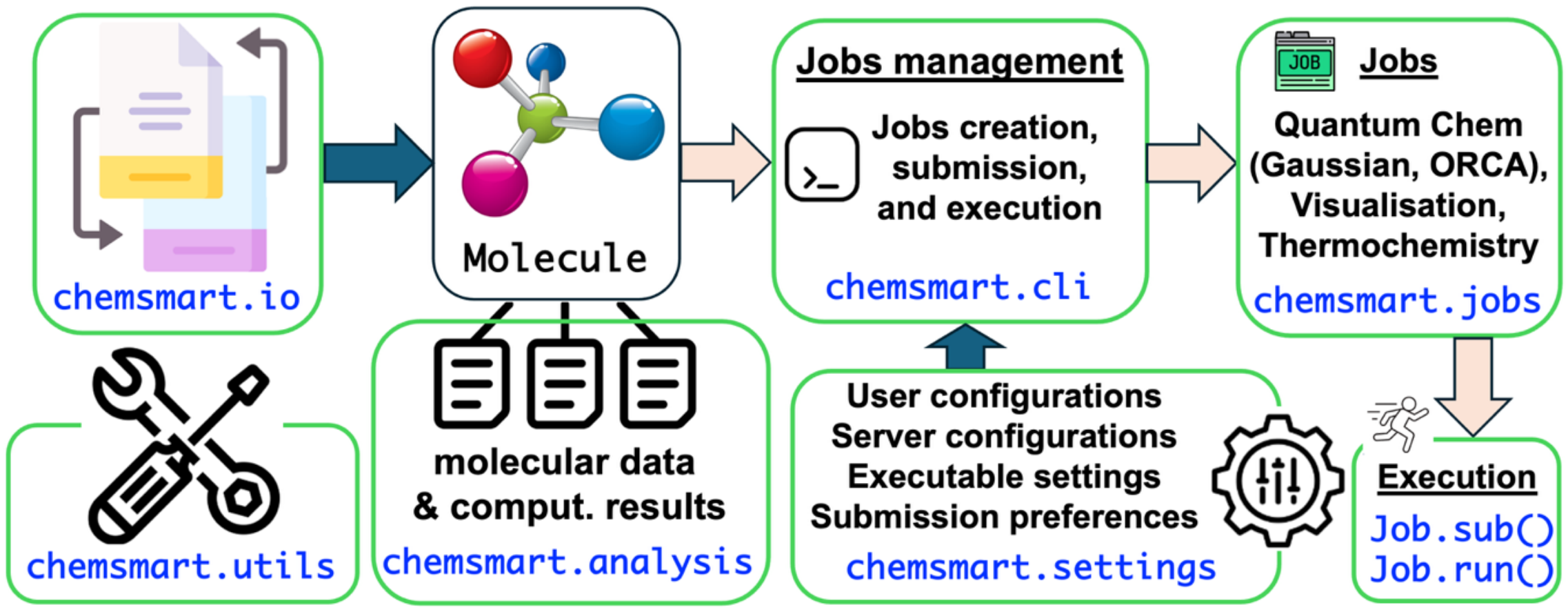

## Abstract

CHEMSMART (Chemistry Simulation and Modeling Automation Toolkit) is an open-source, Python-based framework designed to streamline quantum chemistry workflows for homogeneous catalysis and molecular modeling. By integrating job preparation, submission, execution, results analysis, and visualization, CHEMSMART addresses the inefficiencies of manual workflow management in computational chemistry by ensuring seamless interoperability with quantum chemistry packages and cheminformatics platforms. Its modular architecture supports automated job submission and execution tasks for geometry optimization, transition state searches, thermochemical analysis, and non-covalent interaction plotting, while auxiliary scripts facilitate file conversion, data organization, and electronic structure analysis. Future developments aim to expand compatibility with additional software, incorporate QM/MM and classical MD, and align with FAIR data principles for enhanced reproducibility and data reuse. Available on GitHub, CHEMSMART empowers researchers with a robust, user-friendly platform for efficient and reproducible computational chemistry.

## 1. Introduction

Computational chemistry has become an indispensable tool in modern chemistry across different fields, including drug discovery,[1] materials science,[2] catalysis,[3] and molecular modelling[4]. Homogeneous catalysis, where molecular catalysts and reactants operate in a single phase (often liquid), underpins advances in pharmaceutical synthesis, sustainable chemistry, and energy storage.[5,6] These systems demand precise modelling of solvent effects, transition states, and conformational dynamics, tasks that remain computationally intensive and laborious. By employing computational approaches, researchers can elucidate reaction mechanisms, predict molecular behaviors, and optimize chemical processes, in a wide range of chemistries including radical and photochemistry,[7–11] asymmetric catalysis,[12–23] transition-metal catalysis[24–33] and polymer chemistry modelling[34–38], thereby complementing experimental approaches. Quantum chemistry (QC) software such as Gaussian,[39,40] ORCA,[41–43] and other computational chemistry packages play essential roles in these tasks, providing highly accurate simulations for molecular properties and reactions.

While heterogeneous catalysis and materials modelling benefit from robust automation frameworks such as the Atomic Simulation Environment (ASE)[44] and Pymatgen (Python Materials Genomics)[45], which provide modular interfaces for job management, data handling, and integration into high-throughput workflows, analogous frameworks for molecular systems are far less developed. Some tools partially fill this gap. For instance, *autodE* automates reaction profile generation for both organic and organometallic systems across multiple electronic structure packages;[46] *GoodVibes* simplifies thermochemical analysis by parsing quantum chemical outputs and applying corrections to the rigid-rotor harmonic oscillator approximation, enabling transparent and automated thermochemistry workflows;[47] *AQME* integrates quantum chemistry workflows with cheminformatics for streamlined descriptor extraction.[48] However, each remains specialized to a subset of tasks and a unified, extensible integration for end-to-end molecular quantum chemistry workflows remains lacking.

Driven by the ever-growing computational demands, with application needs in high-throughput screening, multiscale simulations, and machine learning integration, the absence of an integrated workflows management environment has emerged as a critical bottleneck. This is because computational chemistry workflows often encounter significant operational inefficiencies, particularly related to manual job management, resource allocation, and error handling. Manual management is inherently inefficient and error prone. In addition, different job management systems like SLURM, Torque, and other general-purpose schedulers on different high-performance computing (HPC) clusters, will require different submission scripts, complicating their integration into high-throughput and domain-specific workflows. Thus, an integrated workflow package that works from input job preparation to job submission and results analysis will streamline the computational chemistry practices greatly.

Here, we introduce CHEMSMART (Chemistry Simulation and Modeling Automation Toolkit), a comprehensive, open-source Python-based toolkit (GitHub repository https://github.com/xinglong-zhang/chemsmart) designed to unify every step of the quantum chemistry workflow, from input preparation to job execution, results visualization and analysis as well as data organization. Its modular design and adaptable interfaces allow integration with various computing environments and workflows as well as future development with minimal code restructuring. This breadth of scope, combined with its forward-compatible design, positions CHEMSMART as a general-purpose platform for accelerating molecular quantum chemistry workflows while remaining adaptable to future research needs. In this article, we present the design principles and the implementation of CHEMSMART package, emphasizing on its modularity, scalability, and interoperability and demonstrate how CHEMSMART may be installed and used for daily computational chemistry research.

## 2. CHEMSMART Package Design and Implementation

CHEMSMART is structured around a highly modular Python-based architecture. Its object-oriented nature allows for easy adaptation and customization by users. We designed the CHEMSMART package to make it easily extensible, with the possibility of incorporating other software packages without too much additional work. CHEMSMART has been tested to work on different operating systems including Windows, Linux and macOS operating systems.

The toolkit is divided into distinct components responsible for different aspects of computational workflow management (Figure 1). Central to this package is the `Molecule` object, which is a Python class representing the molecular structure. Similar to an ASE `Atoms` object, it contains the construction or initialization parameters reflecting the key information of a molecule. In our case however, the construction parameters are adapted specifically for representing a molecule in a typical homogenous catalysis application. These constructors include the chemical symbols, atomic positions, charge, multiplicity, list of atoms that are frozen or constrained, energy and forces etc. This central `Molecule` object serves as a common object to which files from different chemical simulation packages can be converted, enabling both the storage of computed data and seamless interconversion between different file formats.

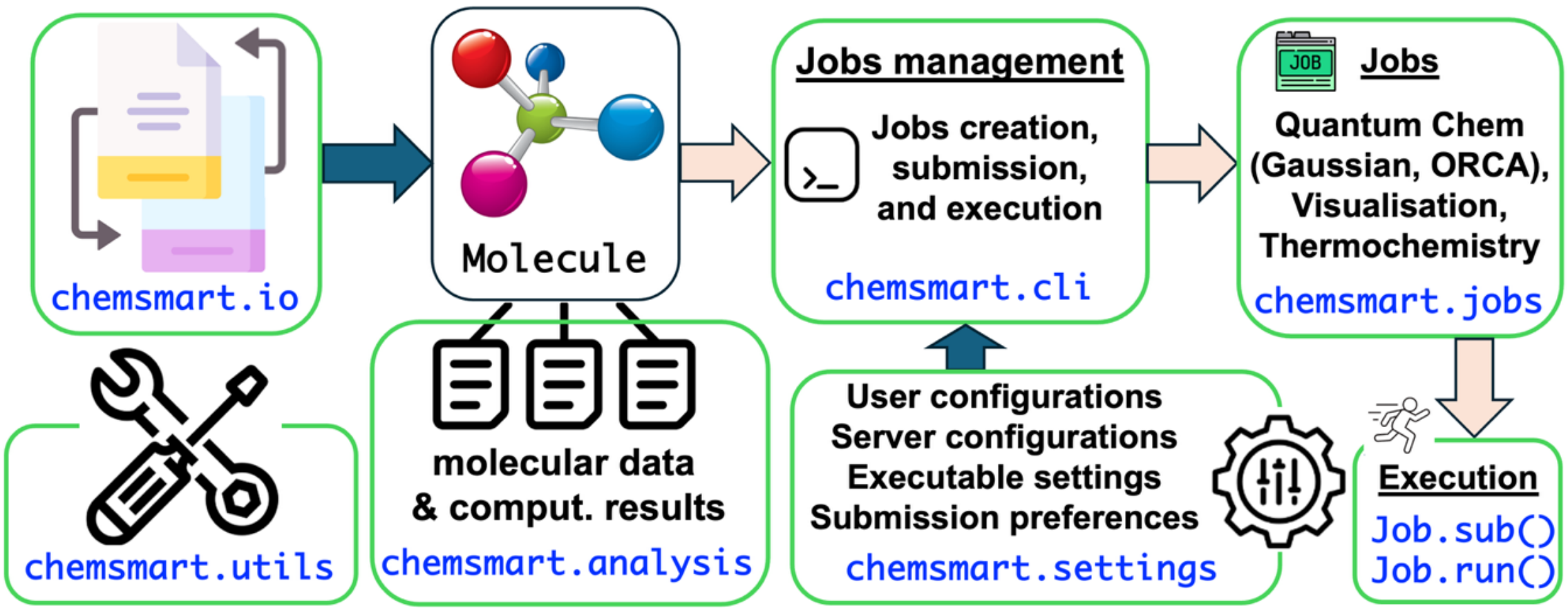

**Figure 1.** Overall design and implementation of the CHEMSMART package.

Beyond the central `Molecule` object, CHEMSMART comprises several key modules that support a streamlined computational workflow. The `chemsmart.io` module handles input and output operations, enabling parsing and generation of chemical file formats from various simulation packages. Complementing this is `chemsmart.utils`, which provides general-purpose tools and utilities to support the broader functionality of the package.

Job creation, submission, and execution are managed through `chemsmart.cli`, a command-line interface designed to facilitate user interaction with minimal overhead. It connects directly with the `chemsmart.jobs` module, which creates various computational jobs for running quantum chemical calculations (e.g., with Gaussian, ORCA) and associated post-processing tasks such as visualization or thermochemical analysis. Each job has a dedicated method for it to be submitted (via job.sub()) or to be run (via job.run()) (Figure 1).

User-specific settings as well as project settings (e.g., the density functional theory (DFT) functional and basis sets required for simulating specific chemical systems) are handled via the `chemsmart.settings` module, which reads user-customizable configuration preferences such as executable paths, HPC server configurations, and submission parameters. We allow each user the flexibility to have their own project and server configurations without modifying the CHEMSMART codes (or needing to know them). These configurations ensure reproducibility and allow users to tailor the package to different computing environments with ease. See examples later.

Finally, the `chemsmart.analysis` module is responsible for managing molecular data and computational results. It allows users to process and analyze the outputs from quantum chemistry calculations, ensuring that important results such as energies, forces, and thermochemical data are accessible and organized.

Together, these modules form a coherent, extensible framework that simplifies the management of electronic structure workflows in homogenous catalysis research.

## 3. CHEMSMART Installation and Configurations

### 3.1 Installation

The source codes for CHEMSMART are publicly available on GitHub. We recommend managing the CHEMSMART toolkit using the conda environment manager. To install, once the conda (base) environment is activated, one need to first clone the repository into the desired directory. On Unix/Linux terminal or Windows PowerShell, one can run:

```
git clone https://github.com/xinglong-zhang/chemsmart.git
```

Then, we go into the `chemsmart` directory and run the following:

```
make env
```

This will create the conda environment `chemsmart`, which will set up the conda environment for the package. At the end of this environment creation, the terminal will show that CHEMSMART package and conda environment is successfully installed:

```
Successfully built chemsmart
Installing collected packages: chemsmart
Successfully installed chemsmart-0.1.0
done
#
# To activate this environment, use
#
#     $ conda activate chemsmart
#
```

After activating the `chemsmart` environment, we can run

```
make install
```

to install additional package dependencies required by CHEMSMART.

We plan to streamline the installation process in the future such that the installation of package can be achieved via a single command:

```
pip install chemsmart
```

### 3.2 Configuration

Once CHEMSMART is installed, configuration is simple and allows for the project and server settings to be set up properly, so that users can use these settings to run computational chemistry projects that may require different DFT methods on different HPC servers. To configure, we will simply run

```
make configure
```

which will 1) create a folder `~/.chemsmart` containing user-specific project and server settings that allows for customization and modification without changing the CHEMSMART codes; 2) update the CHEMSMART path variables automatically in the user's `~/.bashrc` file; 3) update the path variables for Gaussian, ORCA and NCIPLOT[49] variables via user inputs, where the following prompts *requiring user input* will be shown during the configuration process:

```
Enter the path to the Gaussian g16 folder (or press Enter to skip):
```

```
Enter the path to the ORCA folder (or press Enter to skip):
```

```
Enter the path to the NCIPLOT folder (or press Enter to skip):
```

This completes the configurations for the CHEMSMART package and the software executables for running quantum chemistry simulations (via Gaussian and ORCA program) and non-covalent interaction (NCI) plots (via NCIPLOT program).

To allow for the automatic writing of submission scripts, we configure the HPC servers. In the `~/.chemsmart/server/` folder, template files containing the YAML files for configuring SLURM and Torque/PBS scheduler types are included. Users may need to configure the server properties such as QUEUE_NAME, whose value will be the partition name on some schedulers, NUM_HOURS, which specifies the wall time limit on that server, MEM_GB which specifies the memory of each node, for their specific HPC server. Note that different server YAML files may be added to this folder to specify different partition names, allowing for flexible customization.

For project settings containing the DFT methods, users can specify these in the project YAML files located in the `~/.chemsmart/gaussian/` folder, for use in Gaussian calculations and in the `~/.chemsmart/orca/` folder, for use in ORCA calculations. Suppose we have Gaussian job settings for running a project related to iron catalysis, then, we can save these in the file `~/.chemsmart/gaussian/iron.yaml`, with the following contents:

```
gas:
  # ab_initio: MP2  # uncomment to use hf, mp2, ccsd(t) etc
  functional: mn15
  basis: def2svp
  # semiempirical: pm6  # uncomment to semiempirical method
  # solvent_model: smd  # uncomment to use solvent
  # solvent_id: DiChloroEthane  # uncomment to use solvent
solv:
  functional: mn15
  basis: def2tzvp
  freq: False
  solvent_model: smd
  solvent_id: DiChloroEthane
```

where the conventions that the functional and basis specified under `gas` will be used for gas-phase calculations including geometry optimization and transition state search, whereas those under `solv` will be used for single-point solvent correction. For optimizations in solvent directly, one may specify the solvent model and solvent ID in the specification under `gas`. These specifications will allow CHEMSMART to automatically prepare the required input files using the user-specified DFT methods tailored to specific projects. Instead of specifying DFT functional and basis sets used for calculation, one may also specify ab initio quantum chemistry methods, such as HF, MP2, CCSD(T) (requiring basis sets) etc., or semiempirical methods available in Gaussian, such as AM1, PM3, PM6 or PM7. See our *readthedocs* for more details and information.

## 4. CHEMSMART Use Cases

### 4.1 Automation of quantum computational chemistry jobs

Once the project and server settings have been set up, one can submit quantum computational chemistry from the terminal/command line via a single command. A generic command used for job submission is shown below:

```
chemsmart sub gaussian/orca --project <project> --file <file>
opt/ts/<jobtype>
```

which options (such as `--project` and `--file`) can be abbreviated as:

```
chemsmart sub gaussian/orca -p <project> -f <file> opt/ts/<jobtype>
```

See the Supporting Information on full option names and their abbreviations. Here, the `/` denotes choice between alternatives (e.g., `gaussian` or `orca`), and values in `<  >` denote user-specified placeholders. This will submit the appropriate Gaussian (`chemsmart sub gaussian`) or ORCA (`chemsmart sub orca`) job using the level of theory specified in `<project>.yaml` located in the appropriate user project settings folders.

For example, to submit the Gaussian geometry optimization of an iron-containing system in `.xyz` file, we will run the following command:

```
chemsmart sub gaussian -p iron -f input.xyz -c 0 -m 1 opt
```

The flags `-c` and `-m` specify the charge and multiplicity of the system, respectively, as no such information is contained in the `.xyz` file. If unspecified, the charge and multiplicity values will be read from the supplied file; on the other hand, even if the charge and multiplicity has been given in the input file, one may specify their values via the command line to override those default values (e.g., when the users have optimized a molecule in triplet state but want to use the optimized structure as an input for quintet state, one can simply change the multiplicity at run time via the `-m` flag).

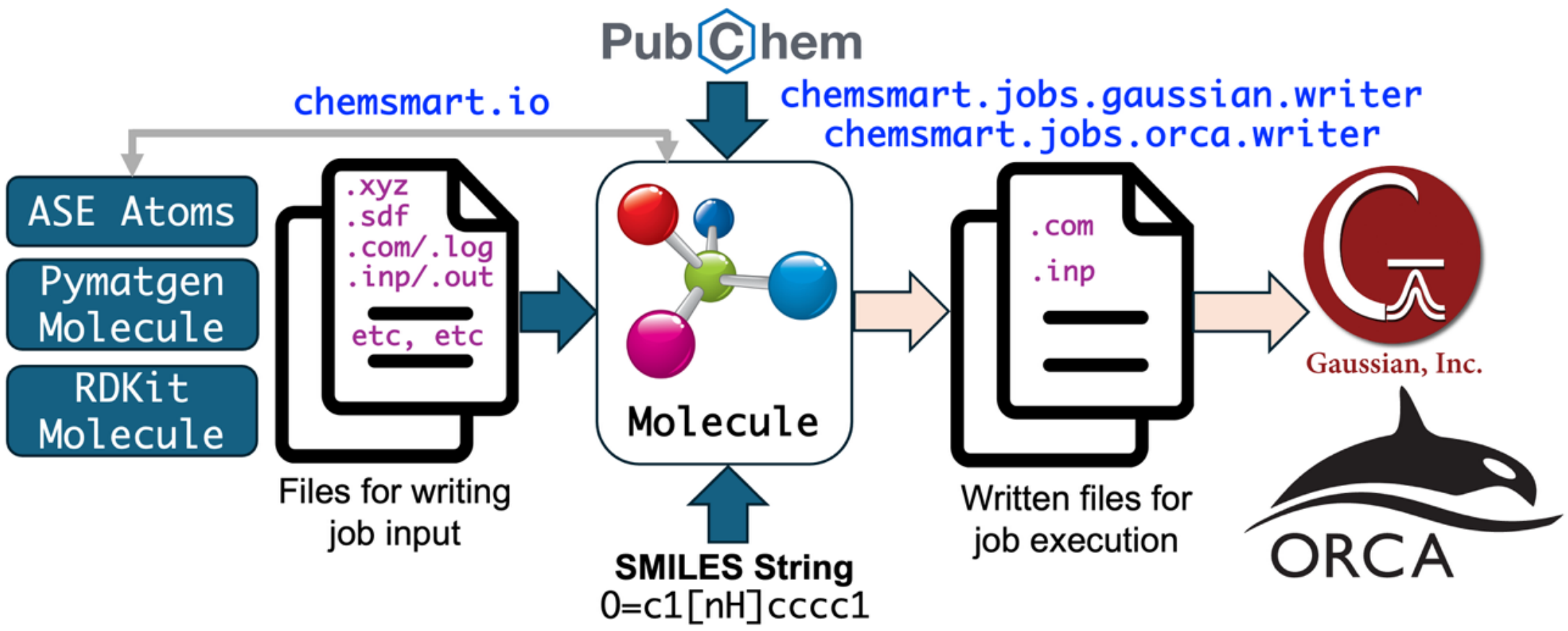

**Figure 2.** Interoperability of the CHEMSMART `Molecule` object across computational chemistry and cheminformatics platforms.

Other types of files may be supplied, including Gaussian `.com`/`.gjf` input file, `.log` output file, ORCA `.inp` input file and `.out` output file. Due to the object-oriented design of our codes, any file type containing a molecular structure may be parsed into the central `Molecule` object, which can be used to write the appropriate input file for the chosen software. This enables seamless interoperability and cross-compatibility between different packages. Thus, one may supply an ORCA file to CHEMSMART to create Gaussian input file and vice versa. Furthermore, input files can be created directly by supplying either the PubChem ID or the SMILES representation of the molecule of interest (Figure 2). For example,

```
chemsmart sub gaussian -p iron -P 10422 -l azetidine -c 0 -m 1 opt
```

will fetch the molecule with PubChem ID 10422 and create an input file named `azetidine.com` (via the label flag `-l azetidine`) with charge 0 and spin multiplicity 1 for Gaussian optimization. The SMILES string of the `azetidine` molecule can be equivalently used, for example, instead of using `-P 10422`, one may use `-P C1CNC1` to submit the same job.

This functionality is particularly advantageous for high-throughput quantum chemistry workflows, where a list of PubChem IDs or SMILES strings can be batch-processed to automatically generate and submit quantum chemistry jobs, thereby eliminating manual file preparation and significantly accelerating large-scale screening campaigns.

The `chemsmart.io` module also allows the interconversion of our `Molecule` object to and from the ASE `Atoms` object, Pymatgen `Molecule` object and RDKit `Mol` object (Figure 2), thus allowing for seamless interoperability across different computational chemistry and cheminformatics ecosystems (allowing access to properties associated with these objects). This enables structures generated or optimized in CHEMSMART to be readily transferred into atomistic simulation workflows (ASE), materials informatics pipelines (Pymatgen), or cheminformatics and machine learning applications (RDKit), and vice versa.

To run other job types, one simply changes the `opt` command into the appropriate command for that job. For example,

```
chemsmart sub gaussian -p iron -f structure.com modred -c [[1,2],[3,4,5]]
```

will run Gaussian modredudant job (`modred` command), where the bond distance between atoms numbered 1 and 2 and the bond angle between atoms 3, 4 and 5 will be frozen.

If this resulting `structure_modred.log` has the desired imaginary frequency, one may use this output file to run the transition state search job (`ts` command) directly:

```
chemsmart sub gaussian -p iron -f structure_modred.log ts
```

We can also run Gaussian potential energy surface scan job (`scan` command):

```
chemsmart sub gaussian -p iron -f structure.com scan -c [1,2] -s -0.1 -n 20
```

where the scanned coordinate (via `-c` flag) here is bond distance between atoms numbered 1 and 2, with a step size of -0.1 Å (via stepsize `-s` flag) with a total of 20 steps (via num-steps `-n` flag).

Suppose a potential energy surface scan was done, producing the PES scan profile as shown below, generated from `structure_scan.log` file, one may use this directly to create input file for TS search via:

```
chemsmart sub gaussian -p iron -f structure_scan.log -i 18 ts
```

where the `-i 18` instructs CHEMSMART to create the input file for TS search using structure at point 18 of the `structure_scan.log` (structure number 3 in Figure 3). Note that this index specification works for any input file that has multiple structures, allowing users to freely select the structure from the supplied file for writing the required input for subsequent quantum chemistry calculations.

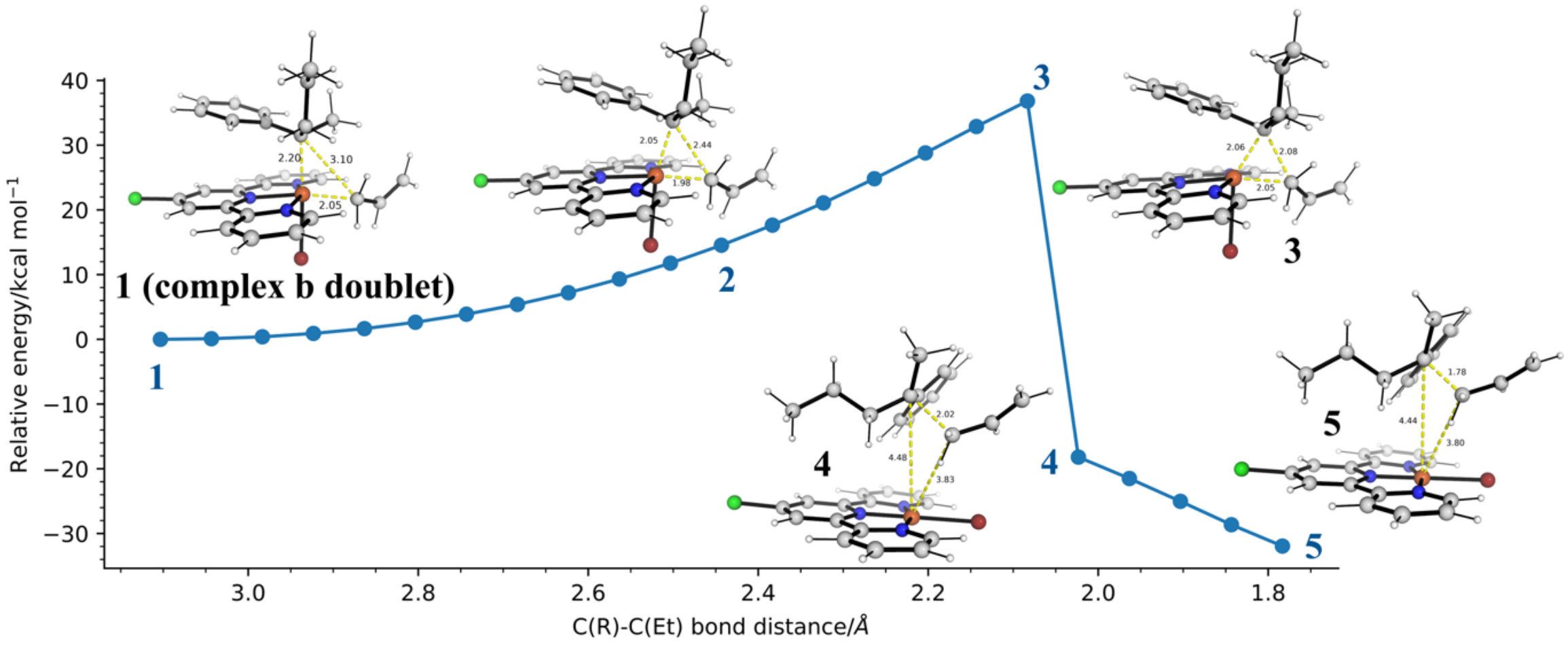

**Figure 3.** Potential energy surface (PES) scan profile generated from `structure_scan.log`, showing key intermediate structures (1–5) along the C(R)–C(Et) bond distance coordinate. Structures from any point in the scan (e.g., structure 3 at index 18) can be directly selected in CHEMSMART program to generate input file for subsequent transition-state search (or any other) job.

Additional customization of each job type is also supported. For instance, optimization-specific parameters such as `maxstep=N` or `maxcycles=N` can be modified using the `-o` flag (--other-opt-options), while other route-string keywords such as `scf=qc`, `nosymm`, or any additional

Gaussian directives, can be appended using the `-r` flag (--additional-route-parameters). For example, the following command

```
chemsmart sub gaussian -p iron -f structure_scan.log -i 18 -o maxstep=5 -
r scf=qc ts
```

will perform TS search with a reduced maximum step size (`maxstep=5`) using the self-consistent field (SCF) procedure with quadratic convergence (`scf=qc`).

### 4.2 Available Gaussian job types for quantum chemistry calculations

Other than the typical Gaussian jobs such as `opt`, `modred`, `ts`, `irc`, `scan`, `sp` and `td` jobs, other specialized job types are available.

#### 4.2.1 Further optimization of CREST conformers or structures from molecular dynamics (MD) trajectories

After a user has performed a *CREST* conformational search,[50,51] using GFN2-xTB[52–54] method, the user may proceed to use the conformers from *CREST* for downstream DFT calculations (such as optimization, modredundant, transition state search, PES scan etc.) directly. For example, if the user wishes to optimize 10 lowest xTB-energy conformers (from the `crest_conformers.xyz` output from *CREST* run) at DFT level, they may simply run

```
chemsmart sub gaussian -p iron -f crest_conformers.xyz -c 0 -m 1 crest -j
opt -N 10
```

where the flag `-N` specifies the number of lowest xTB-energy conformers to optimize (via job type specification (`-j opt`)).

A closely related functionality is the ability to analyze trajectory files (.traj) generated from classical molecular dynamics (MD) simulations and use them as input for quantum chemical refinement. CHEMSMART enables direct submission of density functional theory (DFT) calculations from such trajectories via the command line. For example:

```
chemsmart sub gaussian -p iron -f md.traj -c 0 -m 1 traj -j opt -x 0.1 -g
rmsd -N 10
```

will run DFT calculations with functional and basis sets defined in `iron.yaml` project settings for geometry optimization (`-j opt`). The `-x 0.1` flag specifies that only the final 0.1 proportion of frames from the MD trajectory are selected, and `-g rmsd` will apply a grouping strategy using RMSD to cluster the structures, while `-N 10` restricts the calculation to structures from ten different clusters. This allows efficient sampling of representative configurations from the MD run and subsequent optimization at the DFT level, bridging classical and quantum simulations within a single automated workflow.

### 4.2.2 Distortion-interaction / Activation strain analysis

If a user wishes to run the distortion-interaction model[55,56]/activation strain [56–59] (DI/AS) model to understand the steric and electronic factors influencing the transition state barriers,[60] they may first run the IRC to connect the TS to the reactant. The DI/AS decomposition can then be carried out along selected points of the IRC trajectory using the command:

```
chemsmart sub gaussian -p iron -f ircf.log dias -i 1-20,24,30
```

where `-i` specifies the number of atoms belong to one component of the TS; the other component will be determined by CHEMSMART automatically. At each selected IRC point, the full molecular structure and its two separate fragments are evaluated at the DFT single-point level, enabling automated downstream DI/AS analysis (which is also automated). Additional options are available: for example, `-n 3` will define the number of points or intervals (every 3 points) to sample along the IRC, as well as `-c1 0 -m1 1 -c2 1 -m2 1`, that can be used to specify the charge (`c1, c2`) and spin multiplicity (`m1, m2`) of each fragment, which allows the calculation of fragments that are of different charges (e.g., decomposing a positively charged molecule into a positively charged fragment and a neutral fragment). Once the calculations are done, the distortion and interaction energies at each point will be calculated by calling the script `plot_dias.py` (see discussion on scripts) located in CHEMSMART:

```
plot_dias.py -p gaussian -a 5 -b 7
```

where `-a` and `-b` specifies the atom numbers of the bond distance connecting the two fragments to plot on *x*-axis; `-p` specifies that the output file for plotting is produced from Gaussian (if this DI/AS analysis has been done in ORCA software, then the required value will be `-p orca` instead).

This workflow has been employed in understanding organic[14,21] and organometallic[61] catalysis, where an example of resultant DI/AS plot is given below:

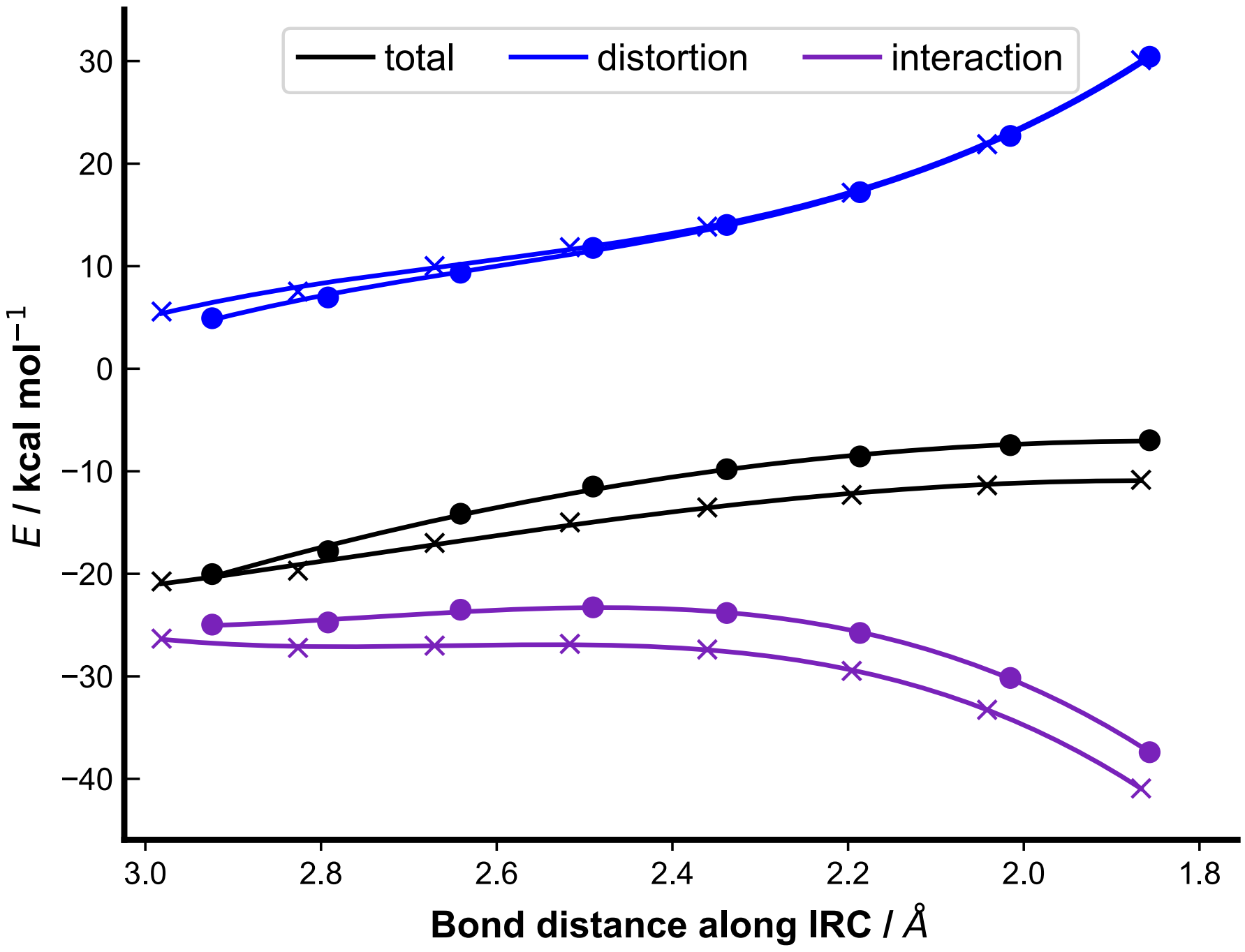

**Figure 4.** An example of the distortion-interaction or activation strain (DI/AS) analyses applied to the IRC paths along two competing transition states marked separately with cross markers vs full circle markers.

The user may also perform DI/AS decomposition using the transition state geometry alone, without using points along the IRC from the reactant to TS, as seen in reference[15]. One may simply specify this via:

```
chemsmart sub gaussian -p iron -f ts.log dias -i 1-20,24,30 -m ts
```

where `-m ts` specifies that only the TS geometry is used (the default mode is `-m irc` as discussed previously).

### 4.2.3 Gaussian --Link1-- job

CHEMSMART also has the option of directly submitting a job requiring the Gaussian --Link1-- in the input file. For example, in the study of openshell singlet systems, symmetry-broken guess wavefunctions are required (via the Gaussian keyword `guess=mix`, which mixes the HOMO and the LUMO to destroy the spatial and spin symmetries). This is typically coupled with the Gaussian keyword `stable=opt`, which checks the stability of the mixed wavefunction before continuation. Users may submit the link job (for `opt`, `modred`, `ts`, `scan`, `sp` etc) via the following command:

```
chemsmart sub gaussian -p iron -f input.xyz -c 0 -m 1 link -j modred -c
[[1,2],[3,4,5],[6,7,8,9]]
```

where the available options for the link job types via `-j` flag can take `opt`, `modred`, `ts`, `scan`, `sp` as values. This will create a Gaussian input file (and submit and run it) where the first route section will be the `stable=opt guess=mix` computation, followed by the appropriate --Link1-- route for the subsequent job. The options to subsequent job types (`opt`, `modred`, `ts`, `scan`, `sp`) are the same as those without Link1 (modredundant would require specification of coordinates to freeze and PES scan would require coordinates to scan, number of steps to scan and the step size to scan).

#### 4.2.4 Miscellaneous jobs

Additional jobs may be submitted via CHEMSMART using a single command in a similar fashion. For example, a non-covalent interaction (NCI) analysis can be initiated with a single command:

```
chemsmart sub gaussian -p iron -f ts.log nci
```

which will automatically create a Gaussian input file (`ts_nci.com`) and execute the job to generate the required `ts_nci.wfn` wavefunction file. Once the `.wfn` file is generated, the NCI plots can be obtained directly by running the following command:

```
chemsmart sub nciplot -f ts_nci.wfn
```

This command submits the NCIPLOT job and produces the `dens.cube` and `grad.cube` files necessary for visualizing the NCI surfaces. Thus, CHEMSMART integrates the generation of wavefunction files using Gaussian software and their subsequent analysis with NCIPLOT program into a streamlined, automated workflow requiring only two commands.

Users may also run NCI plots using promolecular density, which approximates the true electron density of a molecule by summing up independently defined, spherical atomic densities, by simply supplying the `.xyz` file (or any other file such as Gaussian `.log` file or ORCA `.out` file, or PubChem ID or SMILES strings) of the molecule(s) to the command line. For example, the command

```
chemsmart sub nciplot -f very_large_molecule.xyz
```

will submit the job, run NCIPLOT program, and generate the corresponding `dens.cube` and `grad.cube` files using the molecular geometry contained in `very_large_molecule.xyz` file.

Other jobs include restrained electrostatic potential (RESP) fitting (via `resp` command) which fits the classical electrostatics of a molecule a to a quantum-mechanically calculated molecular electrostatic potential (ESP), typically atomic charges; and Wiberg bond index/order calculations in Gaussian (via `wbi` command). See our *readthedocs* for further details.

#### 4.2.5 User-defined jobs and user-prepared Gaussian input files

CHEMSMART also allows users to submit any Gaussian job that falls outside the predefined job types encapsulated within the package. This is achieved through the `userjob` command, where the user explicitly specifies the Gaussian route string and, if needed, any additional append information. For example:

```
chemsmart sub gaussian -p iron -f structure.xyz -c 0 -m 1 userjob -r “opt freq mn15 def2svp” -a “B 1 2 F\nA 3 4 5 F”
```

In this case, the `-r` flag provides the custom route string (`opt freq mn15 def2svp`), while the `-a` flag appends additional information (`B 1 2 F` and `A 3 4 5 F`, one per line) to the end of the coordinate section.

More generally, users may also prepare the Gaussian `.com` or `.gjf` file that they wish to run and submit it via CHEMSMART, which will run it directly, via the `com` subcommand.

For example:

```
chemsmart sub gaussian -p iron -f user_input.gjf com
```

will run the input file supplied by the user, `user_input.gjf`, directly using Gaussian program.

This functionality offers maximum flexibility, enabling users to define and execute arbitrary Gaussian jobs directly from the command line while still benefiting from CHEMSMART's streamlined submission and job management system (no submission script writing is required).

### 4.3 Available ORCA job types for quantum chemistry calculations

Standard ORCA jobs such as geometry optimization (`opt`), modredundant (`modred`), transition state search (`ts`), intrinsic reaction coordinate (`irc`), PES scan (`scan`), and single point (`sp`) calculations can be similarly submitted using the same commands as previously, but replacing `gaussian` by `orca` command, for example,

```
chemsmart sub orca -p iron -f input.xyz -c 1 -m 1 opt
```

will optimize the structure in `input.xyz` with a charge of +1 and multiplicity of 1 using ORCA program.

Given any type of file, CHEMSMART reads in the `Molecule` object, and then use that information to write the input file, it is therefore possible to directly read in e.g., a Gaussian output file and prepare the input file for ORCA calculation. For example,

```
chemsmart sub orca -p dlpno_ccsd_t -f opt.log sp
```

will take the optimized structure in the Gaussian output `opt.log` and write the ORCA single point input using DLPNO-CCSD(T) settings specified in the project YAML file located at user home directory `~/.chemsmart/orca/dlpno_ccsd_t.yaml`. See our *readthedocs* for the detailed examples.

To allow maximum flexibility, CHEMSMART also allows for the submission of a user-prepared ORCA input file directly, similar to the direct submission and execution of a user-prepared Gaussian input file discussed in Section 4.2.5. In this case, users may type a single command,

```
chemsmart sub orca -p iron -f user_input.inp inp
```

to submit and run file supplied by the user, `user_input.inp`, directly using ORCA program (via `inp` command).

### 4.4 Thermochemistry analysis

Once a quantum chemistry job is completed by external software (Gaussian or ORCA), the thermochemistry can be obtained by a single command via the CHEMSMART package. Our chemsmart.analysis.thermochemistry.py (Figure 1) has a `Thermochemistry` class that takes in an output file and processes the data contained in it. A central `Molecule` object containing all key properties (derivable based on molecular geometry alone), such as molecular mass, center of mass, (average) moments of inertia, rotational temperatures, rotational symmetry number, etc., required in the evaluation of the molecular partition functions, can be created from this output file. These values can be calculated from the molecular geometry alone and negates the need to read in from the output file. This approach circumvents potential parsing issues that can arise when programs such as Gaussian output values in fixed-width formats: if the numerical value exceeds the allocated number of digits, Gaussian replaces it with a placeholder such as ****, rather than truncating the number. Such formatting makes direct extraction of the true value from the output file impossible. By calculating these quantities from the molecular structure itself, CHEMSMART avoids these limitations entirely and ensures robust, reproducible thermochemical analysis.

Note that our `Thermochemistry` class includes a Boolean flag, `--use-weighted-mass` (defaults to False, not using), which allows users to choose between using the most abundant isotopic mass (as reported by Gaussian outputs) or the natural abundance-weighted molecular mass. We consider the latter more appropriate, as molecules in practice consist of atoms occurring at their natural isotopic distribution. Accordingly, this weighted mass should be used in the calculation of molecular partition functions and derived thermochemical properties, even though quantum chemical calculations are performed on a single molecular structure using most abundant elements.

Additional thermochemistry information, such as energy values, spin multiplicities and vibrational frequencies may be read from the output file to compute the final thermochemistry. Our modular approach enables that the thermochemistry can be calculated from any type of output files containing an optimized molecular geometry, energy values, spin multiplicities and vibrational frequencies, which can be parsed by `chemsmart.io` module.

To obtain the thermochemistry results from a given output file, one may simply run

```
chemsmart run thermochemistry -f output_file.log -T 298.15 -p 1.0
```

Here, we used `chemsmart run` instead of `chemsmart sub`, as the thermochemistry calculation job is lightweight and may be run directly without creating a job to be submitted to HPC compute node for execution. The evaluation temperature and pressure are specified via the `-T` and `-p` flags, with the value given in Kelvin and atm, respectively. At the specified temperature and pressure, all molecular partition functions, including translational, rotational, vibrational, and electronic, are recalculated (instead of being read from the output file) to prevent double-counting corrections when thermodynamic properties are read from an output file that was generated using a non-default temperature and/or pressure (e.g., in Gaussian input file, by including `temperature=330.15` and/or `pressure=1.5` in the route string).

To apply quasi-harmonic approximations by Grimme[62]/Truhlar[63] to the entropy and/or by Head-Gordon[64] to the enthalpy, one can apply the flags `-csg, --cutoff-entropy-grimme`/`-cst`, `--cutoff-entropy-truhlar` and/or `-ch, --cutoff-enthalpy`, respectively.

Other standard options for thermochemistry calculations, such as temperature, pressure, concentration etc. are also available, one may run

```
chemsmart run thermochemistry --help
```

to get all the help messages and the options available to `thermochemistry` evaluation.

To obtain the thermochemistry results of all the output files in a directory, one can run

```
chemsmart run thermochemistry -d /path/to/directory -t log
```

where `-d` specifies the directory in which to compute the thermochemistry for all output files and `-t` specifies the file type (`log` for Gaussian outputs and `out` for ORCA outputs) to be used.

One may also run a Boltzmann weighted average of thermochemistry properties by calling

```
chemsmart run thermochemistry -d /path/to/directory -t log boltzmann
```

which will compute the thermochemistry for all output files in the specified directory and apply Boltzmann weighting to all the thermochemistry values to output the final Boltzmann-weighted averages.

**4.5 Visualization and molecular analysis**

Our CHEMSMART package integrates open-source PyMOL program[65] for structural and molecular visualization. A PyMOL session file (`.pse`) can be generated automatically to visualize a variety of results with a single command, including molecular structures (`visualize` command), IRC movie connecting reactant and product through the TS (`irc` command), molecular orbitals (`mo` command), NCI plots (`nci` command), spin densities (`spin` command). For example, a molecular structure can be visualized by running:

```
chemsmart run mol -f <structure_file_to_visualize> -i <index> visualize
```

Here, `<structure_file_to_visualize>` may be any file type that can be parsed into a CHEMSMART `Molecule` object (Figure 2). By default, the last structure in the file is selected for visualization, but any structure can be accessed using the `-i` indexing flag (Figure 3), consistent with how indexing is used for input file preparation. Consistent options such as obtaining the structure from PubChem ID or SMILES string for visualization (akin to QC input preparation) are available, for example, one can run

```
chemsmart run mol -P 10422 -l azetidine visualize
```

to fetch the molecule with PubChem ID 10422 and create an `.pse` file containing the `azetidine` molecule for visualization in PyMOL software.

**Illustrated workflow.** Suppose we have a Gaussian guess structure (Figure 5a), saved in file `iron_quintet.com`, that we anticipate undergoing migratory insertion to form C–C bond, we can submit a constrained geometry optimization (modredundant) job by freezing the distances between key atoms (herein between carbon atom #36 and carbon atom #39), using the following command:

```
chemsmart sub gaussian -p iron -f iron_quintet.com modred -c [31,39]
```

When successfully executed, this command produces the output file `iron_quintet_modred.log`. We can then use the last structure of this file to do the TS search, via `ts` command:

```
chemsmart sub gaussian -p iron -f iron_quintet_modred.log ts
```

A successfully located TS is written to `iron_quintet_modred_ts.log`, which can be visualized directly with:

```
chemsmart run mol -f iron_quintet_modred_ts.log visualize -c
[[1,2],[1,9],[1,13],[1,31],[1,36],[1,39],[1,57],[36,39]]
```

Here, the `-c` flag specifies the bonds (`--coordinates`) to display in the visualization (herein bonds between atoms numbered 1 and 2, that between 1 and 9, and so on) (Figure 5b).

Using the file `iron_quintet_modred_ts.log`, we can run Gaussian to generate the wavefunction file (`.wfn`) for NCI plot generation, by running

```
chemsmart sub gaussian -f iron_quintet_modred_ts.log nci
```

which will create the `iron_quintet_modred_ts_nci.wfn`, that can be used for NCI plotting.

Running

```
chemsmart sub nciplot -f iron_quintet_modred_ts_nci.wfn
```

will generate the required `iron_quintet_modred_ts_nci-dens.cube` and `iron_quintet_modred_ts_nci-grad.cube` files for NCI plots visualization. One may also use `iron_quintet_modred_ts.log` (or any file type containing a molecular geometry) directly to generate the `.cube` files, if only promolecular NCI plot is desired:

```
chemsmart sub nciplot -f iron_quintet_modred_ts.log
```

This will generate the `iron_quintet_modred_ts_nci_promolecular-dens.cube` and `iron_quintet_modred_ts_nci_promolecular-grad.cube` files. Once the`.cube` files are generated, we can visualize the NCI surfaces by running:

```
chemsmart run mol -f iron_quintet_modred_ts_nci.log nci
```

or

```
chemsmart run mol -f iron_quintet_modred_ts_nci_promolecular.xyz nci
```

to generate NCI plots from the wavefunction file and from promolecular density, which are shown in Figures 5c and 5d, respectively.

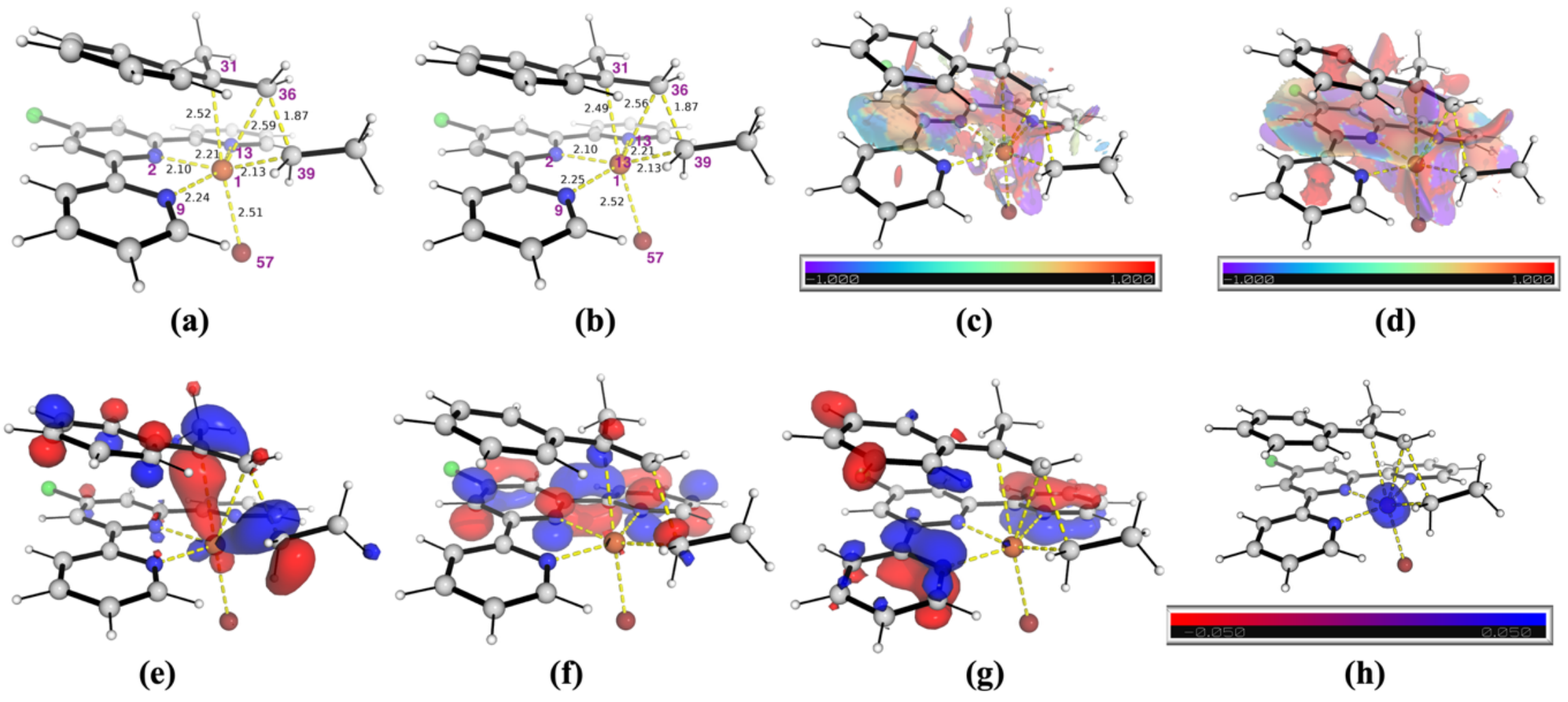

**Figure 5.** Molecular structures (a, b), NCI plots (c, d), molecular orbital plots (e, f, g) and spin density plot (h) generated by running CHEMSMART commands.

Molecular orbitals (MOs) can be visualized directly in CHEMSMART. For example, the highest occupied molecular orbital (HOMO) and lowest unoccupied molecular orbital (LUMO) can be generated with:

```
chemsmart run mol -f iron_quintet_modred_ts.log mo --homo

chemsmart run mol -f iron_quintet_modred_ts.log mo --lumo
```

which will save respectively PyMOL session files, `iron_quintet_modred_ts_HOMO.pse` and `iron_quintet_modred_ts_LUMO.pse`. One may also choose to visualize any MO by explicitly specifying the MO number:

```
chemsmart run mol -f iron_quintet_modred_ts.log mo -n 100
```

Here, MO number 100 is visualized (file `iron_quintet_modred_ts_MO100.pse` is saved). The HOMO, LUMO and MO #100 are shown in Figures 5e, 5f and 5g, respectively.

Spin density plots can likewise be generated with a single command:

```
chemsmart run mol -f iron_quintet_modred_ts.log spin
```

which saves the `iron_quintet_modred_ts_spin.pse` file; the result is shown in Figure 5h.

Note that all these visualizations produce PyMOL `.pse` session files, which can be further manipulated and exported as images within PyMOL. These visualization commands must be executed on a server where the Gaussian program is available (as set in the `~/.chemsmart/server/servername.yaml` files), as the Gaussian `formchk` utility is required to convert checkpoint files into cube files for orbital and spin density visualization.

### 4.6 Auxiliary scripts for streamlining quantum chemistry studies

In addition to the main CHEMSMART modules, we provide a collection of auxiliary Python scripts located in the `chemsmart.scripts` directory. These utilities are designed as lightweight, task-specific tools that address common needs in quantum chemistry workflows, ranging from file handling to post-processing and analysis. By packaging these utilities alongside the main toolkit, CHEMSMART reduces reliance on external *ad hoc* scripts and enables more standardized, reproducible workflows.

#### 4.6.1 File management, conversion and data organization utilities

Several scripts assist with the handling and formatting molecular files and organizing them for publication-ready use. For example, `file_converter.py` (via the `FileConverter` class in `chemsmart.io` module), provides a general-purpose utility for converting structures between commonly used file formats (e.g., `.xyz`, `.sdf`, `.com`, `.mol`, `.inp`); `file_organizer.py` (via the `FileOrganizer` class in `chemsmart.io` module), leverages a simple Excel mapping to reorganize a directory of calculation files into curated folders while renaming runtime filenames to manuscript-ready names, providing a reproducible, one-pass solution for structuring Supporting Information.

| | D | E | F | G | H | I | J | K | L | M | N |
|---|---|---|---|---|---|---|---|---|---|---|---|
| 1 | **remarks** | **folder name** | **name in paper** | **name used during quantum chemical calculations** | | | | **Thermochemistry data** | | | |
| 2 | | | | | | | | | | | |
| 3 | | | | | **E** | **ZPE** | **H** | **T.S** | **T.qh-S** | **G(T)** | **qh-G(T)** |
| 4 | | | | cod2 | -311.34719 | 0.180484 | -311.15515 | 0.047093 | 0.046688 | -311.202241 | -311.2018 |
| 5 | | final_logs | substrate1 | substrate1 | -516.592634 | 0.189856 | -516.38655 | 0.060904 | 0.058183 | -516.447451 | -516.444 |
| 6 | | final_logs | HPPh2 | HPPh2 | -804.018129 | 0.19271 | -803.80916 | 0.062175 | 0.058811 | -803.871333 | -803.8679 |
| 7 | | | | rr_benzp_final | -1307.419377 | 0.401789 | -1306.9857 | 0.091359 | 0.087562 | -1307.07706 | -1307.0732 |
| 8 | | | | NiCOD2_triplet | -2131.053325 | 0.361436 | -2130.6666 | 0.078166 | 0.076308 | -2130.744754 | -2130.7428 |
| 9 | | final_logs | NiCOD2 | NiCOD2 | -2131.111086 | 0.364727 | -2130.7223 | 0.072916 | 0.072045 | -2130.795231 | -2130.794 |
| 10 | | final_logs | difluorophenol | difluorophenol | -505.014972 | 0.089438 | -504.91475 | 0.046048 | 0.046055 | -504.960802 | -504.9608 |
| 11 | | final_logs | sulfonic_acid | sulfonic_acid | -870.896284 | 0.104466 | -870.77902 | 0.051952 | 0.050962 | -870.830968 | -870.8299 |
| 12 | | | | | | | | | | | |
| 13 | | final_logs | PPh2_anion_opt | PPh2_anion_opt | -803.445716 | 0.181289 | -803.24856 | 0.05917 | 0.057394 | -803.30773 | -803.3059 |
| 14 | | final_logs | difluorophenol_anion_ | difluorophenol_anion_opt | -504.453006 | 0.075622 | -504.3669 | 0.045653 | 0.045652 | -504.412553 | -504.4125 |
| 15 | | final_logs | sulfonate_opt | sulfonate_opt | -870.373325 | 0.0927 | -870.26865 | 0.051076 | 0.049715 | -870.319729 | -870.3183 |
| 16 | | final_logs | **INT1** | **NiRRBenzPsub1_binding1_c1_opt** | **-3332.448935** | **0.59562** | **-3331.8039** | **0.129325** | **0.122032** | **-3331.93319** | **-3331.9258** |
| 17 | starts from O coordinated to Ni but optimises to uncoordinated; same as INT1 | | | NiRRBenzPsub1_binding2_opt | -3332.448935 | 0.595623 | -3331.8039 | 0.129317 | 0.122028 | -3331.93318 | -3331.9258 |
| 18 | same as int1 | | | NiRRBenzPsub1_binding2_c1_opt | -3332.448935 | 0.59562 | -3331.8039 | 0.129323 | 0.122033 | -3331.933187 | -3331.9258 |
| 19 | | final_logs | INT1' | xtbopt_NiRRBenzPsub1 | -3332.43917 | 0.595187 | -3331.7941 | 0.132624 | 0.124057 | -3331.926762 | -3331.9181 |
| 20 | | | | NiRRBenzPsub1_opt | -3332.437883 | 0.595315 | -3331.7928 | 0.131736 | 0.123519 | -3331.924558 | -3331.916 |
| 21 | | final_logs | INT1o | NiRRBenzPsub1_c2_opt | -3332.401125 | 0.593297 | -3331.7577 | 0.134029 | 0.124762 | -3331.891701 | -3331.8824 |
| 22 | | | | min | -3332.448935 | 0.593297 | -3331.8039 | 0.129317 | 0.122028 | -3331.93319 | -3331.9258 |
| 23 | | final_logs | INT2A | NiRRBenzPsub1_binding2_acid_inner_Cterm_modred_ts_ircr_rct | -4203.375226 | 0.702856 | -4202.6102 | 0.158396 | 0.14698 | -4202.768612 | -4202.7571 |
| 24 | **inner sphere to INT1 Cterm; major** | final_logs | TS1A_Cterm_c3 | **NiRRBenzPsub1_binding2_acid_inner_Cterm_modred_ts** | **-4203.363995** | **0.698512** | **-4202.6041** | **0.155799** | **0.145043** | **-4202.759909** | **-4202.7491** |
| 25 | | final_logs | INT3A | NiRRBenzPsub1_binding2_acid_inner_Cterm_modred_ts_ircf_prd | -4203.377621 | 0.703734 | -4202.6121 | 0.157735 | 0.14633 | -4202.769841 | -4202.7584 |
| 26 | | final_logs | TS1A_Cterm_c5 | TS1A_Cterm_crest_c30_modred_ts | -4203.364468 | 0.698062 | -4202.6046 | 0.158028 | 0.146691 | -4202.762592 | -4202.7512 |
| 27 | | final_logs | **TS1A_Cterm** | TS1A_Cterm_crest_c1_modred_ts | -4203.371447 | 0.698973 | -4202.6112 | 0.155936 | 0.144921 | -4202.76718 | -4202.7561 |
| 28 | | final_logs | TS1A_Cterm_c4 | TS1A_Cterm_crest_c31_modred_ts | -4203.367748 | 0.698772 | -4202.6076 | 0.157093 | 0.145672 | -4202.764673 | -4202.7532 |
| 29 | | final_logs | TS1A_Cterm_c2 | TS1A_Cterm_crest_c18_modred_ts | -4203.370986 | 0.698922 | -4202.6108 | 0.156075 | 0.144996 | -4202.766889 | -4202.7558 |
| 30 | | | | | | | | | | | |
| 31 | inner sphere to INT1 Cin; minor | final_logs | TS1A_Cin | NiRRBenzPsub1_binding2_acid_inner_Cin_modred_ts | -4203.354998 | 0.698316 | -4202.5949 | 0.156848 | 0.146312 | -4202.751744 | -4202.7412 |
| 32 | | | | | | | | | | | |
| 33 | | | | | | | | | | | |
| 34 | | final_logs | TS1A'_Cterm | acid_INT1prime_Cterm_from_Cin_modred_ts | -4203.350197 | 0.698221 | -4202.5903 | 0.157513 | 0.146613 | -4202.74777 | -4202.736 |
| 35 | | final_logs | TS1A'_Cin_c2 | TS1Aprime_Cin_crest_ts_c8_com | -4203.365757 | 0.698431 | -4202.6057 | 0.156484 | 0.145724 | -4202.762212 | -4202.7514 |
| 36 | | final_logs | TS1A'_Cin | TS1Aprime_Cin_crest_ts_c1_com | -4203.367163 | 0.698385 | -4202.6072 | 0.156694 | 0.145715 | -4202.763935 | -4202.7529 |
| 37 | | final_logs | TS1A'_Cin_c3 | TS1Aprime_Cin_crest_ts_c13_com | -4203.362332 | 0.698082 | -4202.6025 | 0.158433 | 0.146755 | -4202.760905 | -4202.7492 |

**Figure 6.** An Excel file used for data recording during quantum chemistry studies. Column D is used for making remarks on calculations, column E is the designated folder name for collecting final data, column F is the name of the structures used in manuscript, column G is the names of the files used during run time. Column G without column E and F filled are the structures not collected for manuscript. Column H onwards are used for thermochemistry data collection.

An example of such an Excel file is shown in Figure 6, where designated folder names (Column E), final manuscript filenames (Column F), and runtime calculation filenames (Column G) are specified. In this workflow, users can simply call `file_converter.py` to automatically create the designated folder(s), rename files according to the mapping, and transfers them into their appropriate locations. For example, users may run

```
file_organizer.py -f data.xlsx -n sheetname -c E:G -s 2 -r 100 -t log
```

to find Gaussian output files (`-t log`) (or ORCA output files (`-t out`)) in a directory containing `data.xlsx` and output files at various subdirectories and rename the structures accordingly and assemble them in the folder named `final_logs`. These files can then be batch converted to `.xyz` format for downstream use or reporting as supporting information with `file_converter.py`. For example, users may run

```
file_converter.py -d final_logs -t log
```

to convert all the `.log` files in the `final_logs` folder into `.xyz` files (which comment line contains the file base name, the empirical formula, and the final energy in Hartree). The folder containing all `.xyz` files and/or quantum chemical calculation results can be directly used for data reporting in publications,[66–77] for example uploaded to open access data repository such as *zenodo.org*. We strongly advocate for this modern approach as the way forward for data-reporting in quantum chemistry calculations, thus eliminating the outdated practice of including pages of Cartesian coordinates in Supporting Information as static `.pdf` files, which cannot be visualized or reused *directly*. Coordinates in `.pdf` format typically require manual copying and pasting into new files, a process that is error-prone and inconvenient, as the fixed-width layout often breaks the `.xyz` formatting (e.g., copying all $x$-values then y- then z-values). In contrast, `.xyz` files preserve structural information in a machine-readable format, allowing seamless reuse, visualization, and downstream computational analysis.

#### 4.6.2 Electronic structure and property analysis

A suite of scripts is included in CHEMSMART to automate common analyses of electronic structure and molecular properties, thereby reducing the need for manual post-processing and ensuring consistency across studies. These utilities cover a broad range of population analyses, bonding descriptors, and orbital-based methods that are frequently employed in mechanistic and reactivity investigations. For example, `fmo.py` extracts and processes *frontier molecular orbitals* (HOMO and LUMO) and yields the FMO energies, HOMO-LUMO gap, chemical potential, chemical hardness and electrophilicity index, which are essential descriptors for electronic structure, excitation energies, and reactivity trends; `fukui.py` evaluates *Fukui indices*, which quantify the local reactivity of atoms within a molecule and are widely used in conceptual DFT to predict nucleophilic and electrophilic attack sites; `hirshfeld.py` performs *Hirshfeld charge analysis* and offers insights for evaluating atomic charges based on electron density partitioning; `mulliken.py` implements *Mulliken population analysis*, useful for estimating electronic populations and atomic charges; `wbi_analysis.py` computes *Wiberg bond indices*, a measure of bond order derived from the density matrix, which is useful for probing bond strength and electronic delocalization; finally, `cube_operation.py` enables basic manipulations of volumetric data stored in .cube files, including summation, subtraction, and scaling of electron density or orbital maps. These operations support downstream analyses such as electron density difference mapping and visualization of orbital interactions.[78]

Together, these scripts provide a lightweight yet powerful framework for systematically interrogating electronic structures and chemical bonding. By embedding these analyses into automated workflows, CHEMSMART ensures that key descriptors of molecular reactivity and bonding can be obtained in a reproducible and streamlined manner, without requiring users to manually parse and process raw quantum chemistry outputs.

### 4.6.3 Miscellaneous

Other scripts for structure filtering/grouping and automated job submission are also available. We have shown previously the use of `plot_dias.py` in generating plots for DI/AS analysis. We have also `structure_filter.py` which enables the selection or filtering of structures, such as pruning conformer ensembles or extracting representative structures from trajectories. The `submit_jobs.py` provides a lightweight interface for batch job submission (submitting a list of jobs with filenames contained in a `.txt` file, one file per line), complementing the main CHEMSMART sub and run commands. This utility is particularly useful for automating repeated tasks across large datasets.

Together, these auxiliary scripts extend CHEMSMART beyond job preparation and submission, offering a comprehensive toolbox for everyday computational chemistry tasks. They streamline file handling, automate standard analyses (charges, indices, thermochemistry), and integrate advanced workflows such as DI/AS analysis. By bundling these scripts with the main package, CHEMSMART reduces the need for external utilities and consolidates essential tools into a single, user-friendly environment.

### 4.7 Obtaining help

Users may be able to obtain help for available options and commands at each point with the `--help` flag. For example, running

```
chemsmart --help
```

gives available subcommands `config`, `run`, `sub`, `update`, to CHEMSMART; running

```
chemsmart sub --help
```

gives available commands `gaussian`, `mol`, `nciplot`, `orca`, `thermochemistry`, which are available jobs that CHEMSMART can run and automate; running

```
chemsmart sub gaussian --help
```

gives subcommands available to `gaussian`, including typical jobs such as `opt`, `modred`, `ts`, `irc` (IRC, intrinsic reaction coordinates), `scan`, `sp` (single point) and `td` (time-dependent DFT) jobs as well as many others; running

```
chemsmart sub gaussian opt --help
```

will give options available to the `opt` job.

More generally, *every* subcommand in CHEMSMART supports the `--help` option, allowing users to retrieve context-specific guidance interactively at any point in the workflow. Note that the `sub` command will submit the job to the compute node of a HPC cluster for calculation, while the `run` command will run the job (usually lightweight) directly. See the Supporting Information for all the available commands and options.

### 4.8 Using CHEMSMART as a plugin for your own development

In addition to serving as a standalone workflow automation toolkit, CHEMSMART is designed to be modular and can be seamlessly imported as a Python package into user-developed codes, notebooks, or pipelines. This makes it possible to use CHEMSMART in a manner similar to other established libraries such as ASE or Pymatgen.

An illustrated example is shown in Figure 7, where a Jupyter Notebook demonstrates the use of the `Molecule` class from the `chemsmart.io.molecules.structure` module. By reading in a Gaussian output file via `Molecule.from_filepath()` method, the user can directly access a range of key molecular properties, including molecular mass, number of atoms, empirical formula, center of mass, principal moments of inertia, and rotational temperatures. These attributes can then be programmatically retrieved, manipulated, or passed into downstream scripts and analysis workflows without the need for manual parsing of output files.

This plugin-style usage highlights CHEMSMART's interoperability: it not only can act as an end-to-end workflow manager, but also as a flexible building block for custom research applications, data pipelines, and educational demonstrations. By exposing well-structured Python classes and methods, CHEMSMART empowers researchers to incorporate quantum chemistry data and analysis directly into their own computational frameworks.

```
[1]: from chemsmart.io.molecules.structure import Molecule

[2]: mol = Molecule.from_filepath('int_crest_best_singlet.log')

[3]: mol.mass

[3]: 791.5145239959998

[4]: mol.num_atoms

[4]: 93

[5]: mol.empirical_formula

[5]: 'C43H42Cl2FeN2OP2'

[6]: mol.center_of_mass

[6]: array([-0.01304891, -0.00762202,  0.01311489])

[7]: mol.moments_of_inertia

[7]: array([ 7881.92579113,  9401.27701456, 12372.63097552])

[8]: mol.rotational_temperatures

[8]: [0.0030772143444090304, 0.0025799021844017337, 0.0019603247808827402]
```

**Figure 7.** Example Jupyter Notebook session demonstrating the use of CHEMSMART's `Molecule` class to extract structural and thermochemical descriptors from a Gaussian output file.

## 5 Limitations and Further Developments

While CHEMSMART provides an integrated framework for automating quantum chemistry workflows, certain limitations remain that point to directions for further development. At present, CHEMSMART supports Gaussian and ORCA programs, alongside NCIPLOT for non-covalent interaction analysis. Although these packages cover a large fraction of the quantum chemistry community, there is growing demand for integration with additional software, such as Q-Chem,[79] Psi4,[80] NWChem,[81] Molpro,[82] and tight-binding methods (e.g., xTB family). The modular design of CHEMSMART facilitates such extensions, but further development will be required to expand its reach and ensure robust compatibility across diverse simulation engines.

Another major area for future development is the incorporation of multi-scale simulation workflows. Ongoing developments include the automated preparation and running of hybrid quantum mechanics/molecular mechanics (QM/MM) simulations in Gaussian and ORCA software. We also aim to bridge quantum chemistry with classical and machine-learning-based approaches. While CHEMSMART currently focuses on electronic structure methods (Gaussian, ORCA) and related analyses, many other applications such as biomolecular and supramolecular catalysis require coupling quantum calculations with classical molecular dynamics and enhanced sampling. Planned extensions include automated preparation and execution of classical MD workflows (e.g., using GROMACS,[83] PLUMED[84,85]programs) as well as integration with machine learning interatomic potentials (MLIPs). This would allow users to efficiently combine the accuracy of *ab initio* methods with the scalability of classical MD and the predictive efficiency of ML-derived force fields, enabling robust multi-scale simulations for complex systems.

In addition, CHEMSMART promotes modern practices such as exporting `.xyz` and structured result files for deposition in open repositories like *Zenodo* for better reuse and accessibility. However, this currently does not capture the full richness of the quantum chemical data generated from the computationally expensive simulations. Much of these valuable information, such as atomic charges, orbital energies, geometries, energies, forces, vibrational frequencies etc., remains unreported in most computational studies, despite its high utility for machine learning applications. It is worth noting that all these computed quantum chemical properties have already been stored in the `Molecule` object created by CHEMSMART (Figure 7). Ongoing development is looking into data collection and reporting that includes these valuable information. A key goal is to align with the FAIR data principles (Findable, Accessible, Interoperable, Reusable), ensuring that quantum chemical results are stored and disseminated in standardized, machine-readable formats. This will not only facilitate transparent and reproducible computational chemistry but also create a rich, reusable data ecosystem that accelerates progress in areas such as machine learning, data-driven discovery, and automated reaction mechanism exploration.

## 6 Contributing

CHEMSMART is an open-source project, and we welcome contributions from the community to improve and extend its capabilities. Contributions may take the form of new features (e.g., additional job types or software interfaces), bug fixes, improved documentation, or expanded test coverage.

**Getting started.** Developers can fork the CHEMSMART GitHub repository (https://github.com/xinglong-zhang/chemsmart), create a development branch, and work within a conda environment (recommended for consistency). The package should be installed in development mode (make install), and pre-commit hooks (make pre-commit) should be enabled to enforce code style and quality.

**Code style and testing.** CHEMSMART follows a standardized Python style: code formatting with *black* and *isort*, linting with *ruff*, and testing with *pytest*. Contributors are expected to add appropriate tests to ensure full test coverage before submitting pull requests (PRs). Documentation should be updated for any new features or changes in usage.

**Additional dependencies update.** Contributors should run `chemsmart update` (see Supporting Information) to automatically update additional dependencies added in the newly proposed features/codes, to `pyproject.toml` file, which the package uses for the installation of package dependencies.

**Submitting changes.** Contributions should be submitted as pull requests to the main branch. Each PR should include a clear description of the purpose and scope of the changes. Continuous integration (CI) checks will automatically verify formatting, linting, and test coverage.

**Community practices.** To ensure long-term sustainability, CHEMSMART adopts semantic versioning and FAIR principles where applicable. We encourage contributors to help build a robust ecosystem by improving reproducibility, expanding interoperability, and facilitating community adoption.

For detailed developer guidelines, including environment setup, testing, and release procedures, please refer to the README.md and developer documentation in the repository.

## 7 Conclusion

In this work, we introduced CHEMSMART, an open-source toolkit designed to automate and streamline computational chemistry workflows. By integrating job preparation, submission, results parsing, thermochemistry analysis, visualization, and file organization into a coherent and modular framework, CHEMSMART reduces the reliance on *ad hoc* scripting and manual management that often hinders efficiency and reproducibility in quantum chemistry studies.

Central to the package is the `Molecule` object, which enables interoperability between different quantum chemistry engines (currently Gaussian and ORCA) and analysis tools (such as NCIPLOT), as well as compatibility with broader ecosystems like ASE, Pymatgen, and RDKit. The toolkit further provides auxiliary scripts for common tasks such as file conversion, data organization, population analyses, and bond index evaluations, offering users a practical and extensible set of utilities for everyday computational research in molecular science.

To further enhance reproducibility and data accessibility, CHEMSMART advocates for modern data-reporting practices by generating machine-readable `.xyz` files, which preserve structural information for seamless visualization, reuse, and downstream analysis, thereby eliminating the error-prone and outdated practice of embedding Cartesian coordinates in static `.pdf` files.

While CHEMSMART already addresses a wide range of needs in homogeneous catalysis and related areas, limitations remain. At present, support is restricted to a small number of electronic structure packages, and certain advanced tasks still require user oversight. Planned developments include expanding compatibility to additional quantum chemistry and semi-empirical engines (e.g., Q-Chem, Psi4, NWChem, Molpro, xTB), incorporating QM/MM and classical MD workflows (e.g., GROMACS, PLUMED), and integrating machine-learning-based interatomic potentials. Another key direction is aligning CHEMSMART with FAIR data principles, enabling quantum chemistry outputs to be stored in standardized, machine-readable formats for improved accessibility, reproducibility, and reuse.

In summary, CHEMSMART, with its modular and extensible design principles, offers a robust foundation for high-efficiency computational chemistry. By consolidating workflow management, analysis, and data practices into a single platform, it lays the groundwork for more reproducible, scalable, and data-centric high-throughput computational chemistry research.

## Conflicts of interest

There are no conflicts of interest to declare.

## Author Contributions

X.Z. conceptualized the project, directed the overall research, and wrote the manuscript. At the time of writing, X.Z. coded up most of the CHEMSMART package and H. T. coded up the thermochemistry functionalities. H.T., J.L., Z.L., L.W. contributed to code testing and improvements. Z.L. leads the *readthedocs* documentation, with the help from all others. B.W.J.C. provided advice and feedback on code design and code reviews. All authors approved the final manuscript.

## Acknowledgements

This work is supported by the Vice-Chancellor Early Career Professorship Scheme Research Startup Fund (Project Code 4933634) and Research Startup Matching Support (Project Code 5501779) from the Chinese University of Hong Kong.

## Data availability

CHEMSMART package is freely available, under the GPL-3.0 and the LGPL-3.0 licenses, on GitHub at https://github.com/xinglong-zhang/chemsmart. All the files generated, including Gaussian `.chk` and `.log` files, as well as `.cube` and `.pse` files, for the section on **Illustrated workflow** are collected in the folder named `illustrated_workflow`, and uploaded to Zenodo.org at https://zenodo.org/records/16961940 (DOI: 10.5281/zenodo.16961940), under open access. All other additional information has been added in the Supporting Information file.

## References

(1) Sadybekov, A. V.; Katritch, V. Computational Approaches Streamlining Drug Discovery. *Nat. 2023 6167958* **2023**, *616* (7958), 673–685.

(2) Steinhauser, M. O.; Hiermaier, S. A Review of Computational Methods in Materials Science: Examples from Shock-Wave and Polymer Physics. *Int. J. Mol. Sci.* **2009**, *10* (12), 5135–5216.

(3) Thiel, W. Computational Catalysis-Past, Present, and Future. *Angew. Chem. Int. Ed.* **2014**, *53* (33), 8605–8613.

(4) Truhlar, D. G. Molecular Modeling of Complex Chemical Systems. *J. Am. Chem. Soc.*

**2008**, *130* (50), 16824–16827.

(5) Hartwig, J. *Organotransition Metal Chemistry : From Bonding to Catalysis*; University Science Books, U.S., **2010**.

(6) Beller, M.; Renken, A.; van Santen, R. A. *Catalysis: From Principles to Applications*; Wiley, 2012.

(7) Zhang, X.; Paton, R. S. Stereoretention in Styrene Heterodimerisation Promoted by One-Electron Oxidants. *Chem. Sci.* **2020**, *11* (34), 9309–9324.

(8) Liu, D.; Tu, T.; Zhang, T.; Nie, G.; Liao, T.; Ren, S.-C.; Zhang, X.; Chi, Y. R. Photocatalytic Direct Para-Selective C−H Amination of Benzyl Alcohols: Selectivity Independent of Side Substituents. *Angew. Chem. Int. Ed.* **2024**, *63* (43), e202407293.

(9) Zhang, T.; Wang, L.; Peng, X.; Liao, T.; Chen, D.; Tu, T.; Liu, D.; Cheng, Z.; Huang, S.; Ren, S. C.; et al. NHC-Mediated Photocatalytic Para-Selective C–H Acylation of Aryl Alcohols: Regioselectivity Control via Remote Radical Spiro Cyclization. *Sci. China Chem.* **2025**, 1–11.

(10) Wang, G.; Ding, J.; Wu, J. C.; Jin, J.; Zhang, X.; Huang, S.; Ren, S.; Chi, Y. R. Photochemical Dual Radical Coupling of Carboxylates with Alkenes/Heteroarenes via Diradical Equivalents. *J. Am. Chem. Soc.* **2025**, *147* (13), 11368–11377.

(11) Song, T.-T.; Lin, F.; Xu, S.-T.; Zhou, B.-C.; Zhang, L.-M.; Guo, S.-Y.; Zhang, X.; Chen, Q.-A. Divergent Construction of Cyclobutane-Fused Pentacyclic Scaffolds via Double Dearomative Photocycloaddition. *Angew. Chem. Int. Ed.* **2025**, e202505906.

(12) Lv, Y.; Luo, G.; Liu, Q.; Jin, Z.; Zhang, X.; Chi, Y. R. Catalytic Atroposelective Synthesis of Axially Chiral Benzonitriles via Chirality Control during Bond Dissociation and CN Group Formation. *Nat. Commun.* **2022**, *13* (1), 1–9.

(13) Lv, W. X.; Chen, H.; Zhang, X.; Ho, C. C.; Liu, Y.; Wu, S.; Wang, H.; Jin, Z.; Chi, Y. R. Programmable Selective Acylation of Saccharides Mediated by Carbene and Boronic Acid. *Chem* **2022**, *8* (5), 1518–1534.

(14) Yang, X.; Wei, L.; Wu, Y.; Zhou, L.; Zhang, X.; Chi, Y. R. Atroposelective Access to 1,3-Oxazepine-Containing Bridged Biaryls via Carbene-Catalyzed Desymmetrization of Imines. *Angew. Chem. Int. Ed.* **2022**, *62* (1), e202211977.

(15) Wei, L.; Chen, Y.; Zhou, Q.; Wei, Z.; Tu, T.; Ren, S.; Chi, Y. R.; Zhang, X.; Yang, X.

Carbene-Catalyzed Intramolecular Cyclization to Access Inherently Chiral Saddle-Shaped Lactones: Achiral Bases Alternate Product Chirality. *J. Am. Chem. Soc.* **2025**, 147, 34,30747–30756.

(16) Wei, L.; Li, J.; Zhao, Y.; Zhou, Q.; Wei, Z.; Chen, Y.; Zhang, X.; Yang, X. Chiral Phosphoric Acid Catalyzed Asymmetric Hydrolysis of Biaryl Oxazepines for the Synthesis of Axially Chiral Biaryl Amino Phenol Derivatives. *Angew. Chem. Int. Ed.* **2023**, *62* (39), e202306864.

(17) Luo, Z.; Liao, M.; Li, W.; Zhao, S.; Tang, K.; Zheng, P.; Chi, Y. R.; Zhang, X.; Wu, X. Ionic Hydrogen Bond-Assisted Catalytic Construction of Nitrogen Stereogenic Center via Formal Desymmetrization of Remote Diols. *Angew. Chem. Int. Ed.* **2024**, e202404979.

(18) Li, B.; Hu, J.; Liao, M.; Xiong, Q.; Zhang, Y.; Chi, Y. R.; Zhang, X.; Wu, X. Catalyst Control over S(IV)-Stereogenicity via Carbene-Derived Sulfinyl Azolium Intermediates. *J. Am. Chem. Soc.* **2024**, *146*, 25350−25360.

(19) Liu, Y. G.; Zhong, Z.; Tang, Y.; Wang, H.; Vummaleti, S. V. C.; Peng, X.; Peng, P.; Zhang, X.; Chi, Y. R. Carbene-Catalyzed Chirality-Controlled Site-Selective Acylation of Saccharides. *Nat. Commun.* **2025**, *16* (1), 1–10.

(20) Xiong, Q.; Liao, M.; Zhao, S.; Wu, S.; Hong, Y.; Chi, Y. R.; Zhang, X.; Wu, X. Asymmetric Synthesis of S(IV)-Stereogenic Sulfinimidate Esters by Sulfinamide Activation. *Angew. Chem. Int. Ed.* **2025**, *64* (21), e202500170.

(21) Yang, X.; Xie, Y.; Xu, J.; Ren, S.; Mondal, B.; Zhou, L.; Tian, W.; Zhang, X.; Hao, L.; Jin, Z.; et al. Carbene-Catalyzed Activation of Remote Nitrogen Atoms of (Benz)Imidazole-Derived Aldimines for Enantioselective Synthesis of Heterocycles. *Angew. Chem. Int. Ed.* **2021**, *60* (14), 7906–7912.

(22) Song, R.; Liu, Y.; Majhi, P. K.; Ng, P. R.; Hao, L.; Xu, J.; Tian, W.; Zhang, L.; Liu, H.; Zhang, X.; et al. Enantioselective Modification of Sulfonamides and Sulfonamide-Containing Drugs: Via Carbene Organic Catalysis. *Org. Chem. Front.* **2021**, *8* (11), 2413–2419.

(23) Deng, R.; Wu, S.; Mou, C.; Liu, J.; Zheng, P.; Zhang, X.; Chi, Y. R. Carbene-Catalyzed Enantioselective Sulfonylation of Enone Aryl Aldehydes: A New Mode of Breslow Intermediate Oxidation. *J. Am. Chem. Soc.* **2022**, *144* (12), 5441–5449.

(24) Guin, S.; Dolui, P.; Zhang, X.; Paul, S.; Singh, V. K.; Pradhan, S.; Chandrashekar, H. B.; Anjana, S. S.; Paton, R. S.; Maiti, D. Iterative Arylation of Amino Acids and Aliphatic Amines via δ-C($Sp^3$)−H Activation: Experimental and Computational Exploration. *Angew. Chem. Int. Ed.* **2019**, *58* (17), 5633–5638.

(25) Achar, T. K. T. K.; Zhang, X.; Mondal, R.; Shanavas, M. S. S.; Maiti, S.; Maity, S.; Pal, N.; Paton, R. S. R. S.; Maiti, D. Palladium-Catalyzed Directed Meta-Selective C−H Allylation of Arenes: Unactivated Internal Olefins as Allyl Surrogates. *Angew. Chem. Int. Ed.* **2019**, *58* (30), 10353–10360.

(26) Porey, S.; Zhang, X.; Bhowmick, S.; Kumar Singh, V.; Guin, S.; Paton, R. S. R. S.; Maiti, D. Alkyne Linchpin Strategy for Drug:Pharmacophore Conjugation: Experimental and Computational Realization of a Meta-Selective Inverse Sonogashira Coupling. *J. Am. Chem. Soc.* **2020**, *142* (8), 3762–3774.

(27) Das, J.; Ali, W.; Ghosh, A.; Pal, T.; Mandal, A.; Teja, C.; Dutta, S.; Pothikumar, R.; Ge, H.; Zhang, X.; et al. Access to Unsaturated Bicyclic Lactones by Overriding Conventional C(Sp 3)–H Site Selectivity. *Nat. Chem.* **2023**, *15* (11), 1626–1635.

(28) Dutta, U.; Prakash, G.; Devi, K.; Borah, K.; Zhang, X.; Maiti, D. Directing Group Assisted Para-Selective C–H Alkynylation of Unbiased Arenes Enabled by Rhodium Catalysis. *Chem. Sci.* **2023**, *14* (41), 11381–11388.

(29) Porey, S.; Bairagi, Y.; Guin, S.; Zhang, X.; Maiti, D. Nondirected C–H/C–F Coupling for the Synthesis of α-Fluoro Olefinated Arenes. *ACS Catal.* **2023**, *13* (21), 14000–14011.

(30) Bairagi, Y.; Porey, S.; Vummaleti, S. V. C.; Zhang, X.; Lahiri, G. K.; Maiti, D. Synthesis of β-(Hetero)Aryl Ketones via Ligand-Enabled Nondirected C–H Alkylation. *ACS Catal.* **2024**, 15654–15664.

(31) Tan, T.-D.; Serviano, J. M. I.; Luo, X.; Qian, P.-C.; Holland, P. L.; Zhang, X.; Koh, M. J. Congested C($sp^3$)-Rich Architectures Enabled by Iron-Catalysed Conjunctive Alkylation. *Nat. Catal. 2024* **2024**, 1–9.

(32) Tan, T. De; Tee, K. Z.; Luo, X.; Qian, P. C.; Zhang, X.; Koh, M. J. Kinetically Controlled Z-Alkene Synthesis Using Iron-Catalysed Allene Dialkylation. *Nat. Synth.* **2024**, *4* (1), 116–123.

(33) Huang, Z.; Tan, H.; Cui, R.; Hu, Y.; Zhang, S.; Jia, J.; Zhang, X.; Zhang, Q. W.

Regiodivergent Hydrophosphination of Bicyclo[1.1.0]-Butanes under Catalyst Control. *Nat. Commun.* **2025**, *16* (1), 1–10.

(34) Wang, J.; Li, X.; Yi, G.; Teong, S. P.; Chan, S. P.; Zhang, X.; Zhang, Y. Noncrystalline Zeolitic Imidazolate Frameworks Tethered with Ionic Liquids as Catalysts for $CO_2$ Conversion into Cyclic Carbonates. *ACS Appl. Mater. Interfaces* **2024**, *16* (8), 10277–10284.

(35) Zheng, J.; Feng, H.; Zhang, X.; Zheng, J.-W.; Ng, J. K. W.; Wang, S.; Hadjichristidis, N.; Li, Z. Advancing Recyclable Thermosets through C═C/C═N Dynamic Covalent Metathesis Chemistry. *J. Am. Chem. Soc.* **2024**, *146* (31), 21612–21622.

(36) Xiong, S.; Shoshani, M. M.; Zhang, X.; Spinney, H. A.; Nett, A. J.; Henderson, B. S.; Miller, T. F.; Agapie, T. Efficient Copolymerization of Acrylate and Ethylene with Neutral P, O-Chelated Nickel Catalysts: Mechanistic Investigations of Monomer Insertion and Chelate Formation. *J. Am. Chem. Soc.* **2021**, *143* (17), 6516–6527.

(37) Shoshani, M. M.; Xiong, S.; Lawniczak, J. J.; Zhang, X.; Miller, T. F.; Agapie, T. Phosphine-Phenoxide Nickel Catalysts for Ethylene/Acrylate Copolymerization: Olefin Coordination and Complex Isomerization Studies Relevant to the Mechanism of Catalysis. *Organometallics* **2022**, *41* (15), 2119–2131.

(38) Tan, J.; Liu, J.; Zhang, X. Unraveling the Mechanism and Influence of Auxiliary Ligands on the Isomerization of Neutral [P,O]-Chelated Nickel Complexes for Olefin Polymerization. *J. Org. Chem.* **2025**, *90* (5), 2052–2061.

(39) Frisch, M. J. .; Trucks, G. W. .; Schlegel, H. B. .; Scuseria, G. E. .; Robb, M. A. .; Cheeseman, J. R. .; Scalmani, G. .; Barone, V. .; Mennucci, B. .; Petersson, G. A. .; et al. *Gaussian 09*, Revision E.01. Gaussian Inc. Wallingford CT. 2009.

(40) Frisch, M. J. .; Trucks, G. W. .; Schlegel, H. B. .; Scuseria, G. E. .; Robb, M. A. .; Cheeseman, J. R. .; Scalmani, G. .; Barone, V. .; Petersson, G. A. .; Nakatsuji, H. .; et al. Gaussian 16, Revision B.01. 2016.

(41) Neese, F. The ORCA Program System. *Wiley Interdiscip. Rev. Comput. Mol. Sci.* **2012**, *2* (1), 73–78.

(42) Neese, F. Software Update: The ORCA Program System, Version 4.0. *Wiley Interdiscip. Rev. Comput. Mol. Sci.* **2018**, *8* (1), e1327.

(43) Neese, F.; Wennmohs, F.; Becker, U.; Riplinger, C. The ORCA Quantum Chemistry

Program Package. *J. Chem. Phys.* **2020**, *152* (22), 224108.

(44) Blomqvist, J. J.; Dulak, M.; Friis, J.; Hargus, C.; Larsen, A.; Jens, M.; Jakob, B.; Ivano, C.; Rune, C.; Marcin, D.; et al. The Atomic Simulation Environment—a Python Library for Working with Atoms. *J. Phys. Condens. Matter* **2017**, *29* (27), 273002.

(45) Ong, S. P.; Richards, W. D.; Jain, A.; Hautier, G.; Kocher, M.; Cholia, S.; Gunter, D.; Chevrier, V. L.; Persson, K. A.; Ceder, G. Python Materials Genomics (Pymatgen): A Robust, Open-Source Python Library for Materials Analysis. *Comput. Mater. Sci.* **2013**, *68*, 314–319.

(46) Young, T. A.; Silcock, J. J.; Sterling, A. J.; Duarte, F. AutodE: Automated Calculation of Reaction Energy Profiles – Application to Organic and Organometallic Reactions. *Angew. Chem. Int. Ed.* **2021**, *60* (8), 4266–4274.

(47) Luchini, G.; Alegre-Requena, J. V.; Funes-Ardoiz, I.; Paton, R. S. GoodVibes: Automated Thermochemistry for Heterogeneous Computational Chemistry Data. *F1000Research* **2020**, *9*, 291.

(48) Alegre-Requena, J. V.; Sowndarya S. V., S.; Pérez-Soto, R.; Alturaifi, T. M.; Paton, R. S. AQME: Automated Quantum Mechanical Environments for Researchers and Educators. *Wiley Interdiscip. Rev. Comput. Mol. Sci.* **2023**, *13* (5), e1663.

(49) Contreras-García, J.; Johnson, E. R.; Keinan, S.; Chaudret, R.; Piquemal, J. P.; Beratan, D. N.; Yang, W. NCIPLOT: A Program for Plotting Noncovalent Interaction Regions. *J. Chem. Theory Comput.* **2011**, *7* (3), 625–632.

(50) Grimme, S. Exploration of Chemical Compound, Conformer, and Reaction Space with Meta-Dynamics Simulations Based on Tight-Binding Quantum Chemical Calculations. *J. Chem. Theory Comput.* **2019**, *15* (5), 2847–2862.

(51) Pracht, P.; Bohle, F.; Grimme, S. Automated Exploration of the Low-Energy Chemical Space with Fast Quantum Chemical Methods. *Phys. Chem. Chem. Phys.* **2020**, *22* (14), 7169–7192.

(52) Bannwarth, C.; Ehlert, S.; Grimme, S. GFN2-XTB – An Accurate and Broadly Parametrized Self-Consistent Tight-Binding Quantum Chemical Method with Multipole Electrostatics and Density-Dependent Dispersion Contributions. *J. Chem. Theory Comput.* **2019**, *15* (3), 1652–1671.

(53) Grimme, S.; Bannwarth, C.; Shushkov, P. A Robust and Accurate Tight-Binding

Quantum Chemical Method for Structures, Vibrational Frequencies, and Noncovalent Interactions of Large Molecular Systems Parametrized for All Spd-Block Elements (Z = 1-86). *J. Chem. Theory Comput.* **2017**, *13* (5), 1989–2009.

(54) Bannwarth, C.; Caldeweyher, E.; Ehlert, S.; Hansen, A.; Pracht, P.; Seibert, J.; Spicher, S.; Grimme, S. Extended Tight-Binding Quantum Chemistry Methods. *Wiley Interdiscip. Rev. Comput. Mol. Sci.* **2021**, *11* (2), e1493.

(55) Ess, D. H.; Houk, K. N. Distortion/Interaction Energy Control of 1,3-Dipolar Cycloaddition Reactivity. *J. Am. Chem. Soc.* **2007**, *129* (35), 10646–10647.

(56) Bickelhaupt, F. M.; Houk, K. N. Analyzing Reaction Rates with the Distortion/Interaction-Activation Strain Model. *Angew. Chem. Int. Ed.* **2017**, *56* (34), 10070–10086.

(57) Bickelhaupt, F. M. Understanding Reactivity with Kohn-Sham Molecular Orbital Theory: E2-SN2 Mechanistic Spectrum and Other Concepts. *J. Comput. Chem.* **1999**, *20* (1), 114–128.

(58) Fernández, I.; Bickelhaupt, F. M. The Activation Strain Model and Molecular Orbital Theory: Understanding and Designing Chemical Reactions. *Chem. Soc. Rev.* **2014**, *43* (14), 4953–4967.

(59) Wolters, L. P.; Bickelhaupt, F. M. The Activation Strain Model and Molecular Orbital Theory. *Wiley Interdiscip. Rev. Comput. Mol. Sci.* **2015**, *5* (4), 324–343.

(60) Vermeeren, P.; van der Lubbe, S. C. C.; Fonseca Guerra, C.; Bickelhaupt, F. M.; Hamlin, T. A. Understanding Chemical Reactivity Using the Activation Strain Model. *Nat. Protoc.* **2020**, *15* (2), 649–667.

(61) Sinha, S. K.; Panja, S.; Grover, J.; Hazra, P. S.; Pandit, S.; Bairagi, Y.; Zhang, X.; Maiti, D. Dual Ligand Enabled Nondirected C-H Chalcogenation of Arenes and Heteroarenes. *J. Am. Chem. Soc.* **2022**, *144* (27), 12032–12042.

(62) Grimme, S. Supramolecular Binding Thermodynamics by Dispersion-Corrected Density Functional Theory. *Chem.: Eur. J.* **2012**, *18* (32), 9955–9964.

(63) Ribeiro, R. F.; Marenich, A. V.; Cramer, C. J.; Truhlar, D. G. Use of Solution-Phase Vibrational Frequencies in Continuum Models for the Free Energy of Solvation. *J. Phys. Chem. B* **2011**, *115* (49), 14556–14562.

(64) Li, Y. P.; Gomes, J.; Sharada, S. M.; Bell, A. T.; Head-Gordon, M. Improved Force-Field Parameters for QM/MM Simulations of the Energies of Adsorption for Molecules in Zeolites and a Free Rotor Correction to the Rigid Rotor Harmonic Oscillator Model for Adsorption Enthalpies. *J. Phys. Chem. C* **2015**, *119* (4), 1840–1850.

(65) Schrödinger, L. *The PyMOL Molecular Graphics Development Component, Version 1.8*; **2015**.

(66) Zhang, X. Directing Group-Assisted Para-Selective C–H Alkynylation of Unbiased Arenes Enabled by Rhodium Catalysis. Zenodo Dataset. **2023**, https://doi.org/10.5281/zenodo.7585280.

(67) Zhang, X.; Tan, H. Regiodivergent Hydrophosphination of Bicyclo[1.1.0]-Butanes under Catalyst Control. Zenodo Dataset. **2025**, https://doi.org/10.5281/zenodo.15146172.

(68) Zhang, X. Congested C(Sp3)-Rich Architectures by Iron- Catalyzed Conjunctive Alkylation. Zenodo Dataset, **2022**, https://doi.org/10.5281/zenodo.8174459.

(69) Zhang, X. Kinetically Controlled Z-Alkene Synthesis Using Iron-Catalysed Allene Dialkylation. Zenodo Dataset, **2022**, https://doi.org/10.5281/zenodo.11363682.

(70) Zhang, X. Divergent Construction of Cyclobutane-Fused Pentacyclic Scaffolds via Double Dearomative Photocycloaddition. Zenodo Dataset. **2024**, https://doi.org/10.5281/zenodo.14523339.

(71) Zhang, X.; Tan, J.; Liu, J. Unravelling the Mechanism and Influence of Auxiliary Ligands on the Isomerization of Neutral [P,O]-Chelated Nickel Complexes for Olefin Polymerization. Zenodo Dataset, **2024**, https://doi.org/10.5281/zenodo.14186325.

(72) Zhang, X. Asymmetric Synthesis of S(IV)-Stereogenic Sulfinimidate Esters by Sulfinamide Activation. Zenodo Dataset, **2024**, https://doi.org/10.5281/zenodo.13941543.

(73) Zhang, X. NHC-Mediated Photocatalytic Para-Selective C-H Acylation of Aryl Alcohols: Regioselectivity Control via Remote Radical Spiro Cyclization. Zenodo Dataset, **2024**, https://doi.org/10.5281/zenodo.13858725.

(74) Zhang, X. Influence of Keto-Enol Tautomerism in Regulating CO2 Photoreduction Activity in Porous Organic Porphyrinic Photopolymer. Zenodo Dataset, **2024**,

https://doi.org/10.5281/zenodo.11408333.

(75) Zhang, X. Advancing Recyclable Thermosets through Dynamic Covalent Metathesis Based on C=C/C=N Exchange Reactions. Zenodo Dataset, **2024**, https://doi.org/10.5281/zenodo.11280484.

(76) Zhang, X. Non-Crystalline Zeolitic Imidazolate Frameworks Tethered with Ionic Liquids as Catalysts for CO2 Conversion into Cyclic Carbonates. Zenodo Dataset, **2023**, https://doi.org/10.5281/zenodo.10399395.

(77) Zhang, X. Non-Directed C-H/C-F Coupling for the Synthesis of α-Fluoro Olefinated Arenes. Zenodo Dataset, **2023**, https://doi.org/10.5281/zenodo.8374622.

(78) Boruah, A.; Boro, B.; Wang, J.; Paul, R.; Ghosh, R.; Mohapatra, D.; Li, P. Z.; Zhang, X.; Mondal, J. Influence of Keto-Enol Tautomerism in Regulating CO2 Photoreduction Activity in Porous Organic Porphyrinic Photopolymers. *ACS Appl. Mater. Interfaces* **2024**.

(79) Shao, Y.; Gan, Z.; Epifanovsky, E.; Gilbert, A. T. B.; Wormit, M.; Kussmann, J.; Lange, A. W.; Behn, A.; Deng, J.; Feng, X.; et al. Advances in Molecular Quantum Chemistry Contained in the Q-Chem 4 Program Package. *Mol. Phys.* **2015**, *113* (2), 184–215.

(80) Smith, D. G. A.; Burns, L. A.; Simmonett, A. C.; Parrish, R. M.; Schieber, M. C.; Galvelis, R.; Kraus, P.; Kruse, H.; Di Remigio, R.; Alenaizan, A.; et al. P SI4 1.4: Open-Source Software for High-Throughput Quantum Chemistry. *J. Chem. Phys.* **2020**, *152* (18).

(81) Aprà, E.; Bylaska, E. J.; De Jong, W. A.; Govind, N.; Kowalski, K.; Straatsma, T. P.; Valiev, M.; Van Dam, H. J. J.; Alexeev, Y.; Anchell, J.; et al. NWChem: Past, Present, and Future. *J. Chem. Phys.* **2020**, *152* (18).

(82) Werner, H. J.; Knowles, P. J.; Knizia, G.; Manby, F. R.; Schütz, M. Molpro: A General-Purpose Quantum Chemistry Program Package. *Wiley Interdiscip. Rev. Comput. Mol. Sci.* **2012**, *2* (2), 242–253.

(83) Hess, B.; Kutzner, C.; Van Der Spoel, D.; Lindahl, E. GRGMACS 4: Algorithms for Highly Efficient, Load-Balanced, and Scalable Molecular Simulation. *J. Chem. Theory Comput.* **2008**, *4* (3), 435–447.

(84) Bonomi, M.; Branduardi, D.; Bussi, G.; Camilloni, C.; Provasi, D.; Raiteri, P.;

Donadio, D.; Marinelli, F.; Pietrucci, F.; Broglia, R. A.; et al. PLUMED: A Portable Plugin for Free-Energy Calculations with Molecular Dynamics. *Comput. Phys. Commun.* **2009**, *180* (10), 1961–1972.

(85) Tribello, G. A.; Bonomi, M.; Branduardi, D.; Camilloni, C.; Bussi, G. PLUMED 2: New Feathers for an Old Bird. *Comput. Phys. Commun.* **2014**, *185* (2), 604–613.

SUPPORTING INFORMATION

# CHEMSMART: Chemistry Simulation and Modeling Automation Toolkit for High-Efficiency Computational Chemistry Workflows

Xinglong Zhang,[1,*] Huiwen Tan,[1] Jingyi Liu,[1] Zihan Li,[1] Lewen Wang,[2] Benjamin W. J. Chen[3]

[1]*Department of Chemistry, The Chinese University of Hong Kong, Shatin, New Territories, Hong Kong, China*

[2]*Department of Chemistry, Hong Kong Baptist University, Kowloon, Hong Kong, China*

[3]*Institute of High Performance Computing, Agency for Science, Technology and Research (A*STAR), 1 Fusionopolis Way, #16-16, Connexis, Singapore 138632, Singapore*

Email: xinglong.zhang@cuhk.edu.hk

## Table of Contents

Help messages are available for commands and subcommands and available options at each point with the `--help` flag.

## I. Subcommands to chemsmart

Running

```
chemsmart --help
```

gives the following output:

```
Usage: chemsmart [OPTIONS] COMMAND [ARGS]...
Options:
  --verbose
  --help     Show this message and exit.
Commands:
  config
  run
  sub
  update  Manage updates in the chemsmart package.
```

## II. Subcommands to chemsmart config

Running

```
chemsmart config --help
```

gives the following output:

```
Usage: chemsmart config [OPTIONS] COMMAND [ARGS]...
Options:
  --help  Show this message and exit.
Commands:
  gaussian  Configures paths to g16 folder.
  nciplot   Configures paths to NCIPLOT folder.
  orca      Configures paths to ORCA folder.
  server    Configures server settings in ~/.chemsmart/server/*yaml files.
```

Running

```
chemsmart config gaussian --help
```

gives the following output:

```
Usage: chemsmart config gaussian [OPTIONS]
  Configures paths to g16 folder.
  Replaces '~/bin/g16' with the specified folder in YAML files.
  Examples:     chemsmart config gaussian --folder <G16FOLDER>
Options:
  -f, --folder TEXT  Path to the Gaussian g16 folder.  [required]
  --help             Show this message and exit..
```

Running

```
chemsmart config nciplot --help
```

gives the following output:

```
Usage: chemsmart config nciplot [OPTIONS]
  Configures paths to NCIPLOT folder.
  Replaces '~/bin/nciplot' with the specified folder in YAML files.
  Examples:     chemsmart config nciplot --folder <NCIPLOTFOLDER>
Options:
  -f, --folder TEXT  Path to the NCIPLOT folder.  [required]
  --help             Show this message and exit.
```

Running

```
chemsmart config orca --help
```

gives the following output:

```
Usage: chemsmart config orca [OPTIONS]
  Configures paths to ORCA folder.
  Replaces '~/bin/orca' with the specified folder in YAML files.
  Examples:     chemsmart config orca --folder <ORCAFOLDER>
Options:
  -f, --folder TEXT  Path to the ORCA folder.  [required]
  --help             Show this message and exit.
```

Running

```
chemsmart config server --help
```

gives the following output:

```
Usage: chemsmart config server [OPTIONS]

  Configures server settings in ~/.chemsmart/server/*yaml files.

  Add conda env vars after the lines EXTRA_COMMANDS: | # extra commands
to

  activate chemsmart environment in submission script in the *yaml file.

  Examples:     chemsmart config server

Options:

  --help  Show this message and exit.
```

Running

```
chemsmart run --help
```

gives the following output:

```
Usage: chemsmart run [OPTIONS] COMMAND [ARGS]...

Options:

  -s, --server TEXT          Server. If not specified, will try to

                             automatically determine and use the current

                             server.

  -n, --num-cores INTEGER    Number of cores for each job.

  -g, --num-gpus INTEGER     Number of gpus per node. Defaults to number
of

                             GPUs on specified server if None.

  -m, --mem-gb INTEGER       Memory in GBs

  --fake / --no-fake         If true, fake jobrunners will be used.

  --scratch / --no-scratch  Run in scratch mode or without scratch
folder.

  -d, --debug / --no-debug  Turns on debug logging.

  --stream / --no-stream    Turns on logging to stdout.

  --help                    Show this message and exit.


Commands:

  gaussian         CLI for running Gaussian jobs using the chemsmart...

  mol              CLI for running PYMOL visualization jobs using the...

  nciplot          CLI for running NCIPLOT jobs using the chemsmart...

  orca             CLI for running ORCA jobs using the chemsmart
framework.

  thermochemistry  CLI for running thermochemistry jobs using the...
```

Here we observed that the subcommands available to `run` and `sub` are the same, thus, we will use `sub` as the illustration for subcommands going forward.

## III. Subcommands to chemsmart run/sub

### III.1 Subcommands to gaussian

Running

```
chemsmart run gaussian --help
```

gives the following output:

```
Usage: chemsmart run gaussian [OPTIONS] COMMAND [ARGS]...
  CLI for running Gaussian jobs using the chemsmart framework.
Options:
  -p, --project TEXT               Project settings.
  -f, --filename TEXT             filename from which new Gaussian input is
                                   prepared.
  -l, --label TEXT                 write user input filename for the job
                                   (without extension)
  -a, --append-label TEXT          name to be appended to file for the job
  -t, --title TEXT                 Gaussian job title.
  -c, --charge INTEGER             charge of the molecule
  -m, --multiplicity INTEGER       multiplicity of the molecule
  -x, --functional TEXT            New functional to run.
  -b, --basis TEXT                 New basis set to run.
  -s, --semiempirical TEXT         Semiempirical method to run.
  -i, --index TEXT                 Index of molecules to use; 1-based
indices.
                                   Default to the last molecule structure.
                                   1-based index.
  -o, --additional-opt-options TEXT
                                   additional opt options
  -r, --additional-route-parameters TEXT
                                   additional route parameters
  -A, --append-additional-info TEXT
```

```
                                  additional information to be appended
at the
                                  end of the input file. E.g, scrf=read
  -C, --custom-solvent TEXT       additional information to be appended
at the
                                  end of the input file. E.g, scrf=read
  -d, --dieze-tag TEXT            dieze tag for gaussian job; possible
options
                                  include "n", "p", "t" to get "#n",
"#p",
                                  "#t", respectively
  --forces / --no-forces          Whether to calculate forces.
  -P, --pubchem TEXT              Queries structure from PubChem using
name,
                                  smiles, cid and conformer information.
  --help                          Show this message and exit.

Commands:
  com       CLI for running Gaussian input file as is.
  crest     CLI for running Gaussian CREST jobs.
  crestopt  CLI for running Gaussian CREST optimization jobs.
  dias      CLI for running Gaussian DI-AS jobs.
  irc       CLI for running Gaussian IRC jobs.
  link      CLI for running Gaussian link jobs.
  modred    CLI for running Gaussian modred jobs.
  nci       CLI for running Gaussian NCI jobs.
  opt       CLI for optimization calculation for Gaussian.
  resp      CLI for running Gaussian RESP jobs.
  scan      CLI for running Gaussian scan jobs.
  sp        CLI for single point calculation for Gaussian.
  td        CLI for running Gaussian TDDFT jobs.
  traj      CLI for running Gaussian set jobs.
  ts        CLI for transition state calculation for Gaussian.
  userjob   CLI for running Gaussian custom jobs.
  wbi       CLI for running Gaussian WBI jobs.
```

Running com subcommand:

```
chemsmart run gaussian -p test -P 123 -l test com --help
```

gives the following output (here the project settings and molecular structure are required inputs, thus we used dummy `-p test -P 123 -l test`):

```
Usage: chemsmart run gaussian com [OPTIONS]
  CLI for running Gaussian input file as is.
Options:
  -S, --skip-completed / -R, --no-skip-completed
                                  To run completed job again. Use -R to
rerun
                                  completed job.
  --help                          Show this message and exit.
```

Running crest subcommand:

```
chemsmart run gaussian -p test -P 123 -l test crest --help
```

gives the following output:

```
Usage: chemsmart run gaussian crest [OPTIONS]
  CLI for running Gaussian CREST jobs.
Options:
  -S, --skip-completed / -R, --no-skip-completed
                                  To run completed job again. Use -R to
rerun
                                  completed job.
  -j, --jobtype TEXT              Gaussian job type. Options: ["opt",
"ts",
                                  "modred", "scan", "sp"]
  -c, --coordinates TEXT          List of coordinates to be fixed for
modred
                                  or scan job. 1-indexed.
  -s, --step-size TEXT            Step size of coordinates to scan.
  -n, --num-steps TEXT            Step size of coordinates to scan.
-N, --num-confs-to-run INTEGER  Number of conformers to optimize.
  --help                          Show this message and exit.
```

Running `dias` subcommand:

```
chemsmart run gaussian -p test -P 123 -l test dias --help
```

gives the following output:

```
Usage: chemsmart run gaussian dias [OPTIONS]
  CLI for running Gaussian DI-AS jobs.
Options:
  -S, --skip-completed / -R, --no-skip-completed
                                  To run completed job again. Use -R to
rerun
                                  completed job.
  -i, --fragment-indices TEXT     Indices of one fragment for DI-AS
analysis.
                                  [required]
  -n, --every-n-points INTEGER    Every nth points along the IRC file to
                                  prepare for DI-AS analysis.
  -s, --solv / --no-solv          Turn on/off solvent for DI-AS job
                                  calculations.
  -m, --mode [irc|ts]             Mode of DI-AS analysis. Default is IRC.
  -c1, --charge-of-fragment1 INTEGER
                                  Charge of fragment 1.
  -m1, --multiplicity-of-fragment1 INTEGER
                                  Multiplicity of fragment 1.
  -c2, --charge-of-fragment2 INTEGER
                                  Charge of fragment 2.
  -m2, --multiplicity-of-fragment2 INTEGER
                                  Multiplicity of fragment 2.
  --help                          Show this message and exit.
```

Running `irc` subcommand:

```
chemsmart run gaussian -p test -P 123 -l test irc --help
```

gives the following output:

```
Usage: chemsmart run gaussian irc [OPTIONS]
  CLI for running Gaussian IRC jobs.
Options:
  -S, --skip-completed / -R, --no-skip-completed
                                  To run completed job again. Use -R to
rerun
                                  completed job.
  -fl, --flat-irc / --no-flat-irc
                                  whether to run flat irc or not
  -pt, --predictor [LQA|HPC|EulerPC|DVV|Euler]
                                  Type of predictors used for IRC.
Examples
                                  include[HPC, EulerPC, LQA, DVV, Euler].
  -rc, --recorrect [Never|Always|Test]
                                  Recorrection step of HPC and EulerPC
IRCs.
                                  options are: ["Never", "Always",
"Test"].
  -rs, --recalc-step INTEGER      Compute the Hessian analytically every
N
                                  predictor steps or every |N| corrector
steps
                                  if N<0.
  -p, --maxpoints INTEGER         Number of points along reaction path to
                                  examine.
  -c, --maxcycles INTEGER         Maximum number of steps along IRC to
run.
  -s, --stepsize INTEGER          Step size along reaction path, in units
of
                                  0.01 Bohr.
  --help                          Show this message and exit.
```

Running `link` subcommand:

```
chemsmart run gaussian -p test -P 123 -l test link --help
```

gives the following output:

```
Usage: chemsmart run gaussian link [OPTIONS]

  CLI for running Gaussian link jobs.

Options:
  -S, --skip-completed / -R, --no-skip-completed
                                  To run completed job again. Use -R to
rerun
                                  completed job.
  -j, --jobtype TEXT              Gaussian job type. Options: ["opt",
"ts",
                                  "modred", "scan", "sp"]
  -c, --coordinates TEXT          List of coordinates to be fixed for
modred
                                  or scan job. 1-indexed.
  -s, --step-size TEXT            Step size of coordinates to scan.
  -n, --num-steps TEXT            Step size of coordinates to scan.
  -r, --remove-solvent / --no-remove-solvent
                                  Whether to use solvent model in the
job.
                                  Defaults to project settings.
  -sm, --solvent-model TEXT       Solvent model to be used for single
point.
  -si, --solvent-id TEXT          Solvent ID to be used for single point.
  -so, --solvent-options TEXT     Additional solvent options in scrf=()
route.
                                  E.g., `iterative` in
                                  scrf=(smd,water,iterative) viachemsmart
sub
                                  -s xz gaussian -p dnam -f outout.log -a
                                  scrf_iter sp -so iterative
  -st, --stable TEXT              Gaussian stability test. See
                                  https://gaussian.com/stable/ for
options.
                                  Defaults to "stable=opt".
  -g, --guess TEXT                Gaussian guess options. See
                                  https://gaussian.com/guess/ for
options.
                                  Defaults to "guess=mix".
  -r, --route TEXT                Route for link section.
  --help                          Show this message and exit.
```

Running modred subcommand:

```
chemsmart run gaussian -p test -P 123 -l test modred --help
```

gives the following output:

```
Usage: chemsmart run gaussian modred [OPTIONS]
  CLI for running Gaussian modred jobs.
Options:
  -S, --skip-completed / -R, --no-skip-completed
                                  To run completed job again. Use -R to
rerun
                                  completed job.
  -j, --jobtype TEXT              Gaussian job type. Options: ["opt",
"ts",
                                  "modred", "scan", "sp"]
  -c, --coordinates TEXT          List of coordinates to be fixed for
modred
                                  or scan job. 1-indexed.
  -s, --step-size TEXT            Step size of coordinates to scan.
  -n, --num-steps TEXT            Step size of coordinates to scan.
  --help                          Show this message and exit.
```

Running nci subcommand:

```
chemsmart run gaussian -p test -P 123 -l test nci --help
```

gives the following output:

```
Usage: chemsmart run gaussian nci [OPTIONS]
  CLI for running Gaussian NCI jobs.
Options:
  -S, --skip-completed / -R, --no-skip-completed
                                  To run completed job again. Use -R to
rerun
                                  completed job.
  --help                          Show this message and exit.
```

Running opt subcommand:

```
chemsmart run gaussian -p test -P 123 -l test opt --help
```

gives the following output:

```
Usage: chemsmart run gaussian opt [OPTIONS]

  CLI for optimization calculation for Gaussian.

Options:

  -S, --skip-completed / -R, --no-skip-completed
                                  To run completed job again. Use -R to
rerun
                                  completed job.
  -f, --freeze-atoms TEXT         Indices of atoms to freeze for
constrained
                                  optimization.
  --help                          Show this message and exit.
```

Running resp subcommand:

```
chemsmart run gaussian -p test -P 123 -l test resp --help
```

gives the following output:

```
Usage: chemsmart run gaussian resp [OPTIONS]

  CLI for running Gaussian RESP jobs.

Options:

  -S, --skip-completed / -R, --no-skip-completed
                                  To run completed job again. Use -R to
rerun
                                  completed job.
  --help                          Show this message and exit.
```

Running scan subcommand:

```
chemsmart run gaussian -p test -P 123 -l test scan --help
```

gives the following output:

```
Usage: chemsmart run gaussian scan [OPTIONS]

  CLI for running Gaussian scan jobs.

Options:

  -S, --skip-completed / -R, --no-skip-completed
                                  To run completed job again. Use -R to
rerun
                                  completed job.
  -j, --jobtype TEXT              Gaussian job type. Options: ["opt",
"ts",
```

```
                                  "modred", "scan", "sp"]
  -c, --coordinates TEXT          List of coordinates to be fixed for
modred
                                  or scan job. 1-indexed.
  -s, --step-size TEXT            Step size of coordinates to scan.
  -n, --num-steps TEXT            Step size of coordinates to scan.
  --help                          Show this message and exit.
```

Running sp subcommand:

```
chemsmart run gaussian -p test -P 123 -l test sp --help
```

gives the following output:

```
Usage: chemsmart run gaussian sp [OPTIONS]
  CLI for single point calculation for Gaussian.
Options:
  -S, --skip-completed / -R, --no-skip-completed
                                  To run completed job again. Use -R to
rerun
                                  completed job.
  -r, --remove-solvent / --no-remove-solvent
                                  Whether to use solvent model in the
job.
                                  Defaults to project settings.
  -sm, --solvent-model TEXT       Solvent model to be used for single
point.
  -si, --solvent-id TEXT          Solvent ID to be used for single point.
  -so, --solvent-options TEXT     Additional solvent options in scrf=()
route.
                                  E.g., `iterative` in
                                  scrf=(smd,water,iterative) viachemsmart
sub
                                  -s xz gaussian -p dnam -f outout.log -a
                                  scrf_iter sp -so iterative
  --help                          Show this message and exit.
```

Running td subcommand:

```
chemsmart run gaussian -p test -P 123 -l test td --help
```

gives the following output:

```
Usage: chemsmart run gaussian td [OPTIONS]
  CLI for running Gaussian TDDFT jobs.
Options:
  -S, --skip-completed / -R, --no-skip-completed
                                  To run completed job again. Use -R to
rerun
                                  completed job.
  -s, --states [singlets|triplets|50-50]
                                  States for closed-shell singlet
systems.
                                  Options choice =["Singlets",
"Triplets",
                                  "50-50"]
  -r, --root INTEGER              Specifies the “state of interest”. The
                                  default is the first excited state
(N=1).
  -n, --nstates INTEGER           Solve for M states (the default is 3).
If
                                  50-50, this gives the number of each
type of
                                  state to solve (i.e., 3 singlets and 3
                                  triplets).
  -e, --eqsolv TEXT               Whether to perform equilibrium or non-
                                  equilibrium PCM solvation. NonEqSolv is
the
                                  default except for excited state opt
and
                                  when excited state density is requested
                                  (e.g., Density=Current or All).
  --help                          Show this message and exit.
```

Running `traj` subcommand:

```
chemsmart run gaussian -p test -P 123 -l test traj --help
```

gives the following output:

```
Usage: chemsmart run gaussian traj [OPTIONS]

  CLI for running Gaussian set jobs.

Options:
  -S, --skip-completed / -R, --no-skip-completed
                                  To run completed job again. Use -R to
rerun
                                  completed job.
  -j, --jobtype TEXT              Gaussian job type. Options: ["opt",
"ts",
                                  "modred", "scan", "sp"]
  -c, --coordinates TEXT          List of coordinates to be fixed for
modred
                                  or scan job. 1-indexed.
  -s, --step-size TEXT            Step size of coordinates to scan.
  -n, --num-steps TEXT            Step size of coordinates to scan.
  -N, --num-structures-to-run INTEGER
                                  Number of structures from the list of
unique
                                  structures to run the job on.
  -g, --grouping-strategy
[rmsd|rcm|fingerprint|isomorphism|formula|connectivity]
                                  Grouping strategy to use for grouping.
                                  Available options are 'rmsd',
'tanimoto',
                                  'isomorphism', 'formula',
'connectivity'
  -i, --ignore-hydrogens / --no-ignore-hydrogens
                                  Ignore H atoms in the grouping.
  -p, --num-procs INTEGER         Number of processors to use for
grouper.
  -x, --proportion-structures-to-use FLOAT
                                  Proportion of structures from the end
of
                                  trajectory to use.  Values ranges from
0.0 <
                                  x <=1.0. Defaults to 0.1 (last 10% of
                                  structures).
  --help                          Show this message and exit.
```

Running ts command:

```
chemsmart run gaussian -p test -P 123 -l test ts --help
```

gives the following output (here the project settings and molecular structure are required inputs, thus we used dummy -p test -P 123 -l test):

```
Usage: chemsmart run gaussian ts [OPTIONS]
  CLI for transition state calculation for Gaussian.
Options:
  -S, --skip-completed / -R, --no-skip-completed
                                  To run completed job again. Use -R to
rerun
                                  completed job.
  -f, --freeze-atoms TEXT         Indices of atoms to freeze for
constrained
                                  optimization.
  --help                          Show this message and exit.
```

Running userjob subcommand:

```
chemsmart run gaussian -p test -P 123 -l test userjob --help
```

gives the following output:

```
Usage: chemsmart run gaussian userjob [OPTIONS]


  CLI for running Gaussian custom jobs.


Options:
  -S, --skip-completed / -R, --no-skip-completed
                                  To run completed job again. Use -R to
rerun
                                  completed job.
  -r, --route TEXT                user-defined route  [required]
  -a, --append-info TEXT          information to be appended at the end
of the
                                  file
  --help                          Show this message and exit.
```

Running `wbi` subcommand:

```
chemsmart run gaussian -p test -P 123 -l test wbi --help
```

gives the following output:

```
Usage: chemsmart run gaussian wbi [OPTIONS]
  CLI for running Gaussian WBI jobs.
Options:
  -S, --skip-completed / -R, --no-skip-completed
                                  To run completed job again. Use -R to
rerun
                                  completed job.
  --help                          Show this message and exit.
```

### II.2 Subcommands to orca

Running

```
chemsmart run orca --help
```

gives the following output:

```
Usage: chemsmart run orca [OPTIONS] COMMAND [ARGS]…
  CLI for running ORCA jobs using the chemsmart framework.
Options:
  -p, --project TEXT              Project settings.
  -f, --filename TEXT            filename from which new ORCA input is
                                  prepared.
  -l, --label TEXT                write user input filename for the job
                                  (without extension)
  -a, --append-label TEXT         name to be appended to file for the job
  -t, --title TEXT                ORCA job title.
  -c, --charge INTEGER            charge of the molecule
  -m, --multiplicity INTEGER      multiplicity of the molecule
  -A, --ab-initio TEXT            Ab initio method to be used.
  -x, --functional TEXT           New functional to run.
  -D, --dispersion TEXT           Dispersion for DFT functional.
  -b, --basis TEXT                New basis set to run.
  -a, --aux-basis TEXT            Auxiliary basis set.
```

```
  -e, --extrapolation-basis TEXT  Extrapolation basis set.
  -d, --defgrid [defgrid1|defgrid2|defgrid3]
                                  Grid for numerical integration. Choices
are
                                  ['defgrid1', 'defgrid2', 'defgrid3']
  --scf-tol
[NormalSCF|LooseSCF|SloppySCF|StrongSCF|TightSCF|VeryTightSCF|ExtremeSCF]
                                  SCF convergence tolerance.
  --scf-algorithm [GDIIS|DIIS|SOSCF|AutoTRAH]
                                  SCF algorithm to use.
  --scf-maxiter INTEGER           Maximum number of SCF iterations.
  --scf-convergence FLOAT         SCF convergence criterion.
  --dipole / --no-dipole          Dipole moment calculation.
  --quadrupole / --no-quadrupole  Quadrupole moment calculation.
  --mdci-cutoff [loose|normal|tight]
                                  MDCI cutoff. Choices are ['loose',
'normal',
                                  'tight']
  --mdci-density [none|unrelaxed|relaxed]
                                  MDCI density. Choices are ['none',
                                  'unrelaxed', 'relaxed']
  -i, --index TEXT                index of molecule to use; default to
the
                                  last molecule structure.
  -r, --additional-route-parameters TEXT
                                  additional route parameters
  --forces / --no-forces          Forces calculation.
  -P, --pubchem TEXT              Queries structure from PubChem using
name,
                                  smiles, cid and conformer information.
  --help                          Show this message and exit.

Commands:
  inp     Run an ORCA input job as it is.
  irc
  modred
  opt
```

```
  scan
  sp
  ts
```

Running `inp` subcommand:

```
chemsmart run orca -p test -P 123 -l test inp --help
```

gives the following output:

```
Usage: chemsmart run orca inp [OPTIONS]
  Run an ORCA input job as it is. Only requires the file that is to be
run.
Options:
  -S, --skip-completed / -R, --no-skip-completed
                                  To run completed job again. Use -R to
rerun
                                  completed job.
  --help                          Show this message and exit.
```

Running `irc` subcommand:

```
chemsmart run orca -p test -P 123 -l test irc --help
```

gives the following output:

```
Usage: chemsmart run orca irc [OPTIONS]
  CLI for running ORCA IRC jobs.
Options:
  -S, --skip-completed / -R, --no-skip-completed
                                  To run completed job again. Use -R to
rerun
                                  completed job.
  --maxiter INTEGER               Maximum number of iterations.
  -p, --printlevel INTEGER        Print level.
  -d, --direction [both|forward|backward|down]
                                  IRC drirection. Available options:
both,
                                  forward, backward, down.
  -i, --inithess [read|calc_anfreq|calc_numfreq]
```

```
                                  Initial Hessian. Available options:
read,
                                  calc_anfreq, calc_numfreq.
  -f, --hess-filename TEXT       Filename of initial Hessian.
  -m, --hessmode INTEGER          Hessian mode used for the initial
                                  displacement.Default 0.
  -M, --monitor-internals / --no-monitor-internals
                                  Monitor internals to print out up to
three
                                  internal coordinates
   --init-displ [DE|length]        Initial displacement. Available
options: DE,
                                  length. DE for energy difference,
length for
                                  step size.
  --scale-init-displ FLOAT        Step size for initial displacement from
TS.
                                  Default 0.1 a.u.
  --de-init-displ FLOAT           Energy difference for initial
displacement
                                  based on provided Hessian (Default: 2
mEh)
  --follow-coordtype TEXT         Follow coordinate type. Default
cartesian.
                                  The only option.
  --scale-displ-sd FLOAT          Scaling factor for scaling the 1st SD
                                  step.Default to 0.15.
  --adapt-scale-displ / --no-adapt-scale-displ
                                  Modify Scale_Displ_SD when the step
size
                                  becomes smaller or larger.
  --sd-parabolicfit / --no-sd-parabolicfit
                                  Do a parabolic fit for finding an optimal
SD
                                  step length.
  --interpolate-only / --no-interpolate-only
                                  Only allow interpolation for parabolic
fit,
                                  not extrapolation.
```

```
  --do-sd-corr / --no-do-sd-corr  Do SD correction to 1st step.
  --scale-displ-sd-corr FLOAT     Scaling factor for scaling the
correction
                                  step to the SD step. It is multiplied
by the
                                  length of the final 1st SD step.
  --sd-corr-parabolicfit / --no-sd-corr-parabolicfit
                                  Do a parabolic fit for finding an optimal
                                  correction step length.
  --tolrmsg FLOAT                 Tolerance for RMS gradient (a.u.).
Default
                                  5.e-4
  --tolmaxg FLOAT                 Tolerance for maximum gradient (a.u.).
                                  Default 2.e-3
  -I, --internal-modred TEXT      Internal modred. Up to three internal
                                  coordinates can be defined and values
                                  printed.
  --help                          Show this message and exit.
```

Running modred subcommand:

```
chemsmart run orca -p test -P 123 -l test modred --help
```

gives the following output:

```
Usage: chemsmart run orca modred [OPTIONS]
  CLI for running ORCA modred jobs.
Options:
  -S, --skip-completed / -R, --no-skip-completed
                                  To run completed job again. Use -R to
rerun
                                  completed job.
  -j, --jobtype TEXT              ORCA job type. Options: ["opt", "ts",
                                  "modred", "scan", "sp"]
  -c, --coordinates TEXT          List of coordinates to be fixed for
modred
                                  or scan job. 1-indexed.
  -x, --dist-start TEXT           starting distance to scan, in
Angstroms.
```

```
  -y, --dist-end TEXT             ending distance to scan, in Angstroms.
  -n, --num-steps TEXT            Step size of coordinates to scan.
  --help                          Show this message and exit.
```

Running opt subcommand:

```
chemsmart run orca -p test -P 123 -l test opt --help
```

gives the following output:

```
Usage: chemsmart run orca opt [OPTIONS]
  CLI for running ORCA opt jobs.
Options:
  -S, --skip-completed / -R, --no-skip-completed
                                  To run completed job again. Use -R to
rerun
                                  completed job.
  -f, --freeze-atoms TEXT         Indices of atoms to freeze for
constrained
                                  optimization. 1-indexed.
  -i, --invert-constraints / --no-invert-constraints
                                  Invert the constraints for frozen atoms
in
                                  optimization.
  --help                          Show this message and exit.
```

Running scan subcommand:

```
chemsmart run orca -p test -P 123 -l test scan --help
```

gives the following output:

```
Usage: chemsmart run orca scan [OPTIONS]
  CLI for running ORCA scan jobs.
Options:
  -S, --skip-completed / -R, --no-skip-completed
                                  To run completed job again. Use -R to
rerun
                                  completed job.
  -j, --jobtype TEXT              ORCA job type. Options: ["opt", "ts",
                                  "modred", "scan", "sp"]
```

```
  -c, --coordinates TEXT          List of coordinates to be fixed for
modred
                                  or scan job. 1-indexed.
  -x, --dist-start TEXT           starting distance to scan, in
Angstroms.
  -y, --dist-end TEXT             ending distance to scan, in Angstroms.
  -n, --num-steps TEXT            Step size of coordinates to scan.
  --help                          Show this message and exit.
```

Running `sp` subcommand:

```
chemsmart run orca -p test -P 123 -l test sp --help
```

gives the following output:

```
Usage: chemsmart run orca sp [OPTIONS]
  CLI for running ORCA sp jobs.
Options:
  -S, --skip-completed / -R, --no-skip-completed
                                  To run completed job again. Use -R to
rerun
                                  completed job.
  --help                          Show this message and exit.
```

Running `ts` subcommand:

```
chemsmart run orca -p test -P 123 -l test ts --help
```

gives the following output:

```
Usage: chemsmart run orca ts [OPTIONS]
  CLI for running ORCA ts jobs.
Options:
  -S, --skip-completed / -R, --no-skip-completed
                                  To run completed job again. Use -R to
rerun
                                  completed job.
and exit.
  -j, --jobtype TEXT              ORCA job type. Options: ["opt", "ts",
                                  "modred", "scan", "sp"]
```

```
  -c, --coordinates TEXT          List of coordinates to be fixed for
modred
                                  or scan job. 1-indexed.
  -x, --dist-start TEXT           starting distance to scan, in
Angstroms.
  -y, --dist-end TEXT             ending distance to scan, in Angstroms.
  -n, --num-steps TEXT            Step size of coordinates to scan.
  -i, --inhess / --no-inhess      Option to read in Hessian file.
  -f, --inhess-filename TEXT     Filename of Hessian file.
  -h, --hybrid-hess / --no-hybrid-hess
                                  Option to use hybrid Hessian.
  -a, --hybrid-hess-atoms TEXT    List of atoms to use for hybrid
Hessian.
                                  zero-indexed, e.g. [0, 1, 2, 3]
  --numhess / --no-numhess        Option to use numerical Hessian.
  -s, --recalc-hess INTEGER       Number of steps to recalculate Hessian.
  -t, --trust-radius FLOAT        Trust radius for TS optimization.
  -ts, --tssearch-type TEXT       Type of TS search to perform. Options
are
                                  ["optts", "scants"]
  -fs, --full-scan / --no-full-scan
                                  Option to perform a full scan.
  --help                          Show this message and exit.
```

**II.3 Subcommands to thermochemistry**

Running

```
chemsmart run thermochemistry --help
```

gives the following output:

```
Usage: chemsmart run thermochemistry [OPTIONS] COMMAND [ARGS]...


  CLI for running thermochemistry jobs using the chemsmart framework.
This
  command allows you to compute thermochemistry for Gaussian or ORCA
output
  files. `chemsmart run thermochemistry -f udc3_mCF3_monomer_c9.log -f
  udc3_mCF3_monomer_c29.log  -T 298.15` will save results to
  `udc3_mCF3_monomer_c9.dat` and `udc3_mCF3_monomer_c29.dat`. `chemsmart
run
  thermochemistry -d /path/to/directory -t log -T 298.15 -o
  thermochemistry_results.dat` will compute thermochemistry for all
Gaussian
  log files in the specified directory and save to
  `thermochemistry_results.dat`.


Options:
  -d, --directory TEXT            Directory in which to compute
                                  thermochemistry for all files.
  -t, --filetype TEXT            Type of file to calculate thermochemistry
                                  for, if directory is specified.
  -f, --filenames TEXT           Gaussian or ORCA output files for parsing
                                  thermochemistry.
  -cs, --cutoff-entropy FLOAT     Cutoff frequency for entropy in
wavenumbers
  -ch, --cutoff-enthalpy FLOAT    Cutoff frequency for enthalpy in
wavenumbers
  -c, --concentration FLOAT       Concentration in mol/L
  -p, --pressure FLOAT            Pressure in atm.  [default: 1.0]
  -T, --temperature FLOAT         Temperature in Kelvin.  [required]
  -a, --alpha INTEGER             Interpolator exponent used in the
quasi-RRHO
                                  approximation.  [default: 4]
  -w, --weighted                  Use natural abundance weighted masses
(True)
```

```
                                  or use most abundant masses (False).
Default
                                  to False, i.e., use single isotopic
mass.
  -u, --energy-units [hartree|eV|kcal/mol|kJ/mol]
                                  Units of energetic values.  [default:
                                  hartree]
  -o, --outputfile TEXT           Output file to save the thermochemistry
                                  results. Defaults to None, which will
save
                                  results to file_basename.dat. If
specified,
                                  it will save all thermochemistry
results to
                                  this file.
  -O, --overwrite                 Overwrite existing output files if they
                                  already exist.
  -i, --check-imaginary-frequencies
                                  Check for imaginary frequencies in the
                                  calculations.  [default: True]
  -S, --skip-completed / -R, --no-skip-completed
                                  To run completed job again. Use -R to
rerun
                                  completed job.
  --help                          Show this message and exit.

Commands:
  boltzmann  Run the Boltzmann weighted averaging for thermochemistry
jobs.
```

Running `boltzmann` subcommand:

```
chemsmart run thermochemistry -d . -t log -T 300 boltzmann --help
```

In a directory (`.`) with Gaussian log files (`-t log`) gives the following output:

```
Usage: chemsmart run thermochemistry boltzmann [OPTIONS]
  Run the Boltzmann weighted averaging for thermochemistry jobs.
Options:
  -S, --skip-completed / -R, --no-skip-completed
                                  To run completed job again. Use -R to
rerun
                                  completed job.
  -w, --energy-type-for-weighting [gibbs|electronic]
                                  Type of energy to use for Boltzmann
                                  weighting.  [default: gibbs]
  --help                          Show this message and exit.
```

**II.4 Subcommands to mol**

Running

```
chemsmart run mol --help
```

gives the following output:

```
Usage: chemsmart run mol [OPTIONS] COMMAND [ARGS]...


  CLI for running PYMOL visualization jobs using the chemsmart framework.
  Example usage: chemsmart run mol -f test.xyz visualize -c
  [[413,409],[413,412],[413,505],[413,507]]


Options:
  -f, --filename TEXT      filename from which new Gaussian input is
prepared.
  -l, --label TEXT          write user input filename for the job (without
                            extension)
  -a, --append-label TEXT  name to be appended to file for the job
  -i, --index TEXT          Index of molecules to use; 1-based indices.
Default
                            to the last molecule structure. 1-based index.
  -P, --pubchem TEXT        Queries structure from PubChem using name,
smiles,
                            cid and conformer information.
  --help                    Show this message and exit.

Commands:
  irc        CLI for generating automatic PyMOL IRC movie and saving
the...
  mo         CLI for generating molecular orbitals (MOs) and saving as...
  movie      CLI for generating automatic PyMOL movie for rotating...
  nci        CLI for generating automatic PyMOL NCI plot and saving as...
  spin       CLI for generating spin density and saving as pse file.
  visualize  CLI for running automatic PyMOL visualization and saving
as...
```

Running `irc` subcommand:

```
chemsmart run mol -P 123 -l test irc --help
```

gives the following output:

```
Usage: chemsmart run mol irc [OPTIONS]

  CLI for generating automatic PyMOL IRC movie and saving the pse file.

  Example usage:     chemsmart run mol irc -r vhr_ox_modred_ts10_ircr.log
-p

  vhr_ox_modred_ts10_ircf.log -c [1,12] -o This makes an IRC movie from

  vhr_ox_modred_ts10_ircr.log and vhr_ox_modred_ts10_ircf.log, with
coordinate

  labels. If the movie mp4 file exists, it will not be overwritten unless
-o

  is specified.     chemsmart run mol irc -a vhr_ox_modred_ts10_irc.log -c

  [1,12] for full irc run, with coordinate labels. This can also be used
for

  any .log to obtain all structures as a movie.

Options:

  -S, --skip-completed / -R, --no-skip-completed

                                  To run completed job again. Use -R to
rerun

                                  completed job.

  -f, --file TEXT                 PyMOL file script or style. If not

                                  specified, defaults to

                                  zhang_group_pymol_style.py.

  -s, --style [pymol|cylview]     PyMOL render style. Choices include
"pymol"

                                  or "cylview", if using
```

```
                                  zhang_group_pymol_style.
  -t, --trace / --no-trace        PyMOL options to ray trace or not.
Default
                                  to True.
  -v, --vdw                       Add Van der Waal surface. Default to
False.
  -q, --quiet                     Run PyMOL in quiet mode. Default to
True.
  --command-line-only / --no-command-line-only
                                  Run PyMOL in command line only. Default
to
                                  True.
  -c, --coordinates TEXT          List of coordinates (bonds, angles and
                                  dihedrals) for labelling. 1-indexed.
  -o, --overwrite                 Overwrite existing files. Default to
False.
  -r, --reactant TEXT             IRC file leading to the reactant side.
  -p, --product TEXT              IRC file leading to the product side.
  -a, --all TEXT                  File containing all structures in the
IRC,
                                  from full IRC run.
  --help                          Show this message and exit.
```

Running mo subcommand:

```
chemsmart run mol -P 123 -l test mo --help
```

gives the following output:

```
Usage: chemsmart run mol mo [OPTIONS]
  CLI for generating molecular orbitals (MOs) and saving as pse file.
Example
  usage:     chemsmart run --debug mol -f phenyldioxazolone.com mo --homo
This
  visualizes the HOMO of phenyldioxazolone.com file and saves as
  phenyldioxazolone_HOMO.pse
Options:
  -S, --skip-completed / -R, --no-skip-completed
                                  To run completed job again. Use -R to
rerun
                                  completed job.
```

```
  -f, --file TEXT                PyMOL file script or style. If not
                                  specified, defaults to
                                  zhang_group_pymol_style.py.
  -s, --style [pymol|cylview]     PyMOL render style. Choices include
"pymol"
                                  or "cylview", if using
                                  zhang_group_pymol_style.
  -t, --trace / --no-trace        PyMOL options to ray trace or not.
Default
                                  to True.
  -v, --vdw                       Add Van der Waal surface. Default to
False.
  -q, --quiet                     Run PyMOL in quiet mode. Default to
True.
  --command-line-only / --no-command-line-only
                                  Run PyMOL in command line only. Default
to
                                  True.
  -c, --coordinates TEXT          List of coordinates (bonds, angles and
                                  dihedrals) for labelling. 1-indexed.
  -n, --number INTEGER            Molecular Orbital number to be
visualized
                                  (e.g., 31 will visualize MO #31).
Default to
                                  None.
  -h, --homo                      Plot the highest occupied molecular
orbital
                                  (HOMO). Default to False.
  -l, --lumo                      Plot the lowest unoccupied molecular
                                  orbitals (LUMO). Default to False.
  --help                          Show this message and exit.
```

Running `movie` subcommand:

```
chemsmart run mol -P 123 -l test movie --help
```

gives the following output:

```
Usage: chemsmart run mol movie [OPTIONS]

  CLI for generating automatic PyMOL movie for rotating molecule and
saving as

  pse file. Example usage:     chemsmart run --debug mol -f

  phenyldioxazolone.com movie -v This visualizes phenyldioxazolone.com
file

  and saves as phenyldioxazolone_movie.pse with added van der Waal's
surface

  (-v) automatically. If the movie mp4 file exists, it will not be
overwritten

unless -o is specified.

chemsmart run --debug mol -f vhr_ox_modred_ts10.log visualize -c
[[1,2],[3,4,5],[1,3,4,5],[4,5],[4,6,9]]

  This visualizes vhr_ox_modred_ts10.log file and saves as

  vhr_ox_modred_ts10_visualize.pse and add in additional coordinates
(bonds,

  angles and dihedrals) for labelling.

Options:

  -S, --skip-completed / -R, --no-skip-completed

                                  To run completed job again. Use -R to
rerun

                                  completed job.

  -f, --file TEXT                PyMOL file script or style. If not

                                  specified, defaults to

                                  zhang_group_pymol_style.py.

  -s, --style [pymol|cylview]     PyMOL render style. Choices include
"pymol"

                                  or "cylview", if using

                                  zhang_group_pymol_style.

  -t, --trace / --no-trace        PyMOL options to ray trace or not.
Default

                                  to True.

  -v, --vdw                       Add Van der Waal surface. Default to
False.

  -q, --quiet                     Run PyMOL in quiet mode. Default to
True.

  --command-line-only / --no-command-line-only

                                  Run PyMOL in command line only. Default
to

                                  True.
```

```
  -c, --coordinates TEXT          List of coordinates (bonds, angles and
                                  dihedrals) for labelling. 1-indexed.
  --help                          Show this message and exit.
```

Running nci subcommand:

```
chemsmart run mol -P 123 -l test nci --help
```

gives the following output:

```
Usage: chemsmart run mol nci [OPTIONS]
  CLI for generating automatic PyMOL NCI plot and saving as pse file.
Example
  usage:     chemsmart run --debug mol -f phenyldioxazolone.com visualize
-v
  This visualizes phenyldioxazolone.com file and saves as
  phenyldioxazolone_visualize.pse with added van der Waal's surface (-v)
  automatically.     chemsmart run --debug mol -f vhr_ox_modred_ts10.log
  visualize -c [[1,2],[3,4,5],[1,3,4,5],[4,5],[4,6,9]] This visualizes
  vhr_ox_modred_ts10.log file and saves as
vhr_ox_modred_ts10_visualize.pse
  and add in additional coordinates (bonds, angles and dihedrals) for
  labelling.
Options:
  -S, --skip-completed / -R, --no-skip-completed
                                  To run completed job again. Use -R to
rerun
                                  completed job.
  -f, --file TEXT                PyMOL file script or style. If not
                                  specified, defaults to
                                  zhang_group_pymol_style.py.
  -s, --style [pymol|cylview]     PyMOL render style. Choices include
"pymol"
                                  or "cylview", if using
                                  zhang_group_pymol_style.
  -t, --trace / --no-trace        PyMOL options to ray trace or not.
Default
                                  to True.
  -v, --vdw                       Add Van der Waal surface. Default to
False.
```

```
  -q, --quiet                      Run PyMOL in quiet mode. Default to
True.
  --command-line-only / --no-command-line-only
                                   Run PyMOL in command line only. Default
to
                                   True.
  -c, --coordinates TEXT           List of coordinates (bonds, angles and
                                   dihedrals) for labelling. 1-indexed.
  -i, --isosurface FLOAT           Isosurface value for NCI plot.
Default=0.5
  -r, --color-range FLOAT          Ramp range for NCI plot. Default=1.0
  -b, --binary                     Plot NCI plots with two colors only.
Default
                                   to False.
  --intermediate                   Plot NCI plots with intermediate range
                                   colors. Default to False.
  --help                           Show this message and exit.
```

Running spin subcommand:

```
chemsmart run mol -P 123 -l test spin --help
```

gives the following output:

```
Usage: chemsmart run mol spin [OPTIONS]
  CLI for generating spin density and saving as pse file. Example usage:
  chemsmart run --debug mol -f phenyldioxazolone.log spin This visualizes
  phenyldioxazolone.log file and saves as phenyldioxazolone_spin.pse.
Requires
  phenyldioxazolone.chk be present together with phenyldioxazolone.log

Options:
  -S, --skip-completed / -R, --no-skip-completed
                                   To run completed job again. Use -R to
rerun
                                   completed job.
  -f, --file TEXT                PyMOL file script or style. If not
                                   specified, defaults to
                                   zhang_group_pymol_style.py.
```

```
  -s, --style [pymol|cylview]     PyMOL render style. Choices include
"pymol"
                                  or "cylview", if using
                                  zhang_group_pymol_style.
  -t, --trace / --no-trace        PyMOL options to ray trace or not.
Default
                                  to True.
  -v, --vdw                       Add Van der Waal surface. Default to
False.
  -q, --quiet                     Run PyMOL in quiet mode. Default to
True.
  --command-line-only / --no-command-line-only
                                  Run PyMOL in command line only. Default
to
                                  True.
  -c, --coordinates TEXT          List of coordinates (bonds, angles and
                                  dihedrals) for labelling. 1-indexed.
  --help                          Show this message and exit.
```

Running `visualize` subcommand:

```
chemsmart run mol -P 123 -l test visualize --help
```

gives the following output:

```
Usage: chemsmart run mol visualize [OPTIONS]
  CLI for running automatic PyMOL visualization and saving as pse file.
  Example usage:     chemsmart run --debug mol -f phenyldioxazolone.com
  visualize -v This visualizes phenyldioxazolone.com file and saves as
  phenyldioxazolone_visualize.pse with added van der Waal's surface (-v)
  automatically.     chemsmart run --debug mol -f vhr_ox_modred_ts10.log
  visualize -c [[1,2],[3,4,5],[1,3,4,5],[4,5],[4,6,9]] This visualizes
  vhr_ox_modred_ts10.log file and saves as
vhr_ox_modred_ts10_visualize.pse
  and add in additional coordinates (bonds, angles and dihedrals) for
  labelling.
Options:
  -S, --skip-completed / -R, --no-skip-completed
                                  To run completed job again. Use -R to
rerun
                                  completed job.
```

```
  -f, --file TEXT                PyMOL file script or style. If not
                                  specified, defaults to
                                  zhang_group_pymol_style.py.
  -s, --style [pymol|cylview]     PyMOL render style. Choices include
"pymol"
                                  or "cylview", if using
                                  zhang_group_pymol_style.
  -t, --trace / --no-trace        PyMOL options to ray trace or not.
Default
                                  to True.
  -v, --vdw                       Add Van der Waal surface. Default to
False.
  -q, --quiet                     Run PyMOL in quiet mode. Default to
True.
  --command-line-only / --no-command-line-only
                                  Run PyMOL in command line only. Default
to
                                  True.
  -c, --coordinates TEXT          List of coordinates (bonds, angles and
                                  dihedrals) for labelling. 1-indexed.
  --help                          Show this message and exit.
```

**II.5 Subcommands to nciplot**

Running

```
chemsmart run nciplot --help
```

gives the following output:

```
Usage: chemsmart run nciplot [OPTIONS]
  CLI for running NCIPLOT jobs using the chemsmart framework. Example
usage:
  chemsmart run nciplot -f test.xyz -f test2.xyz -l nci_test --fragments
"{1:
  [1,4,5], 2: [3,4,5]}"
Options:
```

```
  -S, --skip-completed / -R, --no-skip-completed
                                  To run completed job again. Use -R to
rerun
                                  completed job.
  -f, --filenames TEXT            Input files for the NCIPLOT job. Can be
                                  specified multiple times.
  -l, --label TEXT                Label for the NCIPLOT job, used to name
                                  output files.
  -r, --rthres FLOAT              r distance along a cubic box. This
extends
                                  the grid to include a larger region
around
                                  the molecule for capturing NCIs that
extend
                                  further out.
  --ligand-file-number INTEGER    Ligand file number, corresponds to the
nth
                                  file of the input.
  --ligand-radius FLOAT           Radius of interaction from ligand file
n.
  -rp, --radius-positions TEXT    Positions around which interactions are
                                  represented. Accepts strings in the
form of
                                  'x,y,z' or as a tuple (x, y, z).
  -rr, --radius-r FLOAT           Radius from which interactions are
                                  represented.
  -i1, --intercut1 FLOAT          Cutoff 1, r1, for intermolecularity.
  -i2, --intercut2 FLOAT          Cutoff 2, r2, for intermolecularity.
  --increments TEXT               Increments along the x, y, z directions
of
                                  the cube in Å. The default is set to
0.1,
                                  0.1, 0.1.Accepts strings in the form of
                                  'x,y,z' or as a tuple (x, y, z).
  --fragments TEXT                Fragments in a dictionary-type string.
E.g.,
                                  {1: [1, 2, 3], 2: [4, 5, 6]}
```

```
dat
                                  file.
  -crd, --cutoff-rdg-dat FLOAT    Cutoff for RDG (reduced density
gradient)
                                  used in creating the dat file.
  -cdc, --cutoff-density-cube FLOAT
                                  Cutoff for density used in creating the
cube
                                  file.
  -crc, --cutoff-rdg-cube FLOAT   Cutoff for RDG (reduced density
gradient)
                                  used in creating the cube file.
  --dgrid / --no-dgrid            Turn on Radial grids for promolecular
                                  densities. Default is exponential fits.
                                  Exponential fits are available up to
Z=18
                                  (Ar); radial grids are available up to
Pu
                                  (Z=94).  Defaults to exponential fits
unless
                                  the molecule contains some atom with
Z>18 or
                                  there are charged atoms (cations).
Using
                                  DGRID, only radial grids are used to
                                  calculate promolecular densities.
  --integrate / --no-integrate    Trigger the integration of properties.
  --ranges TEXT                   Ranges for computing properties. A
lower and
                                  upper bounds are required for every
interval
                                  (one per line). Example input:
                                  [[-0.1,-0.02],[-0.02,0.02],[0.02,0.1]]
  --grid-quality [coarse|fine|ultrafine]
                                  Quality of the grid used for NCIPLOT
                                  calculations. Options are 'coarse',
'fine',
                                  or 'ultrafine'.
  -P, --pubchem TEXT              Queries structure from PubChem using
```

```
name,
                                  smiles, cid and conformer information.
  --help                          Show this message and exit.
```

**IV. Subcommands to chemsmart update**

Running

```
chemsmart update --help
```

gives the following output:

```
Usage: chemsmart update [OPTIONS] COMMAND [ARGS]...


  Manage updates in the chemsmart package.


Options:
  --help  Show this message and exit.


Commands:
  deps  Automatically update dependencies in pyproject.toml.
```

Running

```
chemsmart update deps --help
```

gives the following output:

```
Usage: chemsmart update deps [OPTIONS]
  Automatically update dependencies in pyproject.toml.
Options:
  --help  Show this message and exit.
```

**V. chemsmart scripts**

Running

```
cube_operation.py --help
```

gives the following output:

```
Usage: cube_operation.py [OPTIONS]

  CLI for running cube operation using the chemsmart framework.

Options:
  -c1, --cube1 TEXT                Cube file 1.  [required]
  -c2, --cube2 TEXT                Cube file 2.  [required]
  -x, --operation [subtract|add]   Operation type. Defaults to subtract.
  -o, --outputname TEXT            outputname of the operated cube file.
  --help                           Show this message and exit.
```

Running

```
file_converter.py --help
```

gives the following output:

```
Usage: file_converter.py [OPTIONS]

  Script for converting structures in different formats.

Options:
  -d, --directory TEXT             Directory in which to convert files.
  -t, --type TEXT                  Type of file to be converted, if
direcotry
                                   is specified.
  -f, --filename TEXT             Input filename to be converted.
  -o, --output-filetype TEXT      Type of files to convert to, defaults to
                                   .xzy
  -i, --include-intermediate-structures / --no-include-intermediate-
structures
                                   Include intermediate structures in the
                                   conversion.
  --help                           Show this message and exit.
```

Running

```
file_organizer.py --help
```

gives the following output:

```
Usage: file_organizer.py [OPTIONS]


  Script for organizing files for supporting information. Example usage:
  file_organizer.py -f jq.xlsx -n co2 -c B:D -s 2 -r 100


Options:
  -d, --directory TEXT            Directory in which to organize the
files.
  -f, --filename TEXT            Filename of Excel file to use for
                                  organizing.  [required]
  -n, --name TEXT                 Sheet name of Excel file to use for
                                  organizing.  [required]
  -t, --type TEXT                 Type of file to be organized.
  -c, --cols TEXT                 Columns to be used.
  -s, --skip INTEGER              Number of rows to skip.
  -r, --row INTEGER               Number of rows to organize.

  --keep-default-na / --no-keep-default-na
                                  Keep default NA values in Excel file.
  --help                          Show this message and exit.
```

Running

```
fmo.py --help
```

gives the following output:

```
Usage: fmo.py [OPTIONS]
  CLI for running fmo using the chemsmart framework.
Options:
  -f, --filename TEXT       Gaussian or ORCA output file.  [required]
  -u, --unit [eV|kcal/mol]  Unit of FMO energy.
  --help                    Show this message and exit.
(chemsmart) [lwang34@bridges2-login014 scripts]$ fmo.py --help
Usage: fmo.py [OPTIONS]

Options:
  -f, --filename TEXT       Gaussian or ORCA output file.  [required]
  -u, --unit [eV|kcal/mol]  Unit of FMO energy.
  --help                    Show this message and exit.
```

Running

```
fukui.py --help
```

gives the following output:

```
Usage: fukui.py [OPTIONS]
    CLI for running fukui using the chemsmart framework.
Options:
  -n, --neutral-filename TEXT     Gaussian or ORCA output file for the
neutral
                                  system.  [required]
  -c, --radical-cation-filename TEXT
                                  Gaussian or ORCA output file for the
radical
                                  cationic system.
  -a, --radical-anion-filename TEXT
                                  Gaussian or ORCA output file for the
radical
                                  anionic system.
  -m, --mode [mulliken|nbo|hirshfeld|cm5]
                                  Charges to be used for Fukui Indices
                                  calculations.
  --help                          Show this message and exit.essage and
exit.
```

Running

```
hirshfeld.py --help
```

gives the following output:

```
Usage: hirshfeld.py [OPTIONS]

    CLI for running hirshfeld using the chemsmart framework.

Options:
  -f, --filename TEXT    Gaussian or ORCA output file.  [required]
  -n, --numbers INTEGER  Atom numbers from which to obtain Hirshfeld
Charges
                         and Spins. 1-indexed.
  --help                 Show this message and exit.
```

Running

```
mulliken.py --help
```

gives the following output:

```
Usage: mulliken.py [OPTIONS]

    CLI for running mulliken using the chemsmart framework.

Options:
  -f, --filename TEXT    Gaussian or ORCA output file.  [required]
  -n, --numbers INTEGER  Atom numbers from which to obtain Hirshfeld
Charges
                         and Spins. 1-indexed.
  --help                 Show this message and exit.
```

Running

```
plot_dias.py --help
```

gives the following output:

```
Usage: plot_dias.py [OPTIONS]


  Example usage: plot_dias.py -p orca -z  -a 5 -b 7 -r.


Options:
  -f, --folder TEXT
  -p, --program TEXT            Type of program: Gaussian or ORCA output
for
                                plotting.
  -z, --zero / --no-zero        Set reference point all to zero.
  -o, --outputname TEXT         output file name for dias data to be
written
  -e, --extrapolate BOOLEAN
  -r, --reverse / --no-reverse
  -n, --new-length INTEGER
  -k, --k-value INTEGER         Degree of the smoothing spline. Must be 1
<= k
                                <= 5. k = 3 is a cubic spline. Default is
3.
  -a, --atom-number1 INTEGER    Atom number 1 for the bond distance to
plot on
                                x-axis.  [required]

  -b, --atom-number2 INTEGER    Atom number 2 for the bond distance to
plot on
                                x-axis.  [required]
  -x, --ref-file TEXT          Reference file as zero for DIAS plot.
  --help                        Show this message and exit.
```

Running

```
structure_filter.py --help
```

gives the following output:

```
Usage: structure_filter.py [OPTIONS]

   CLI for running structure filter using the chemsmart framework.

Options:
  -d, --directory TEXT            Directory containing files to filter.
                                  [required]
  -t, --type TEXT                 Type of files to filter.
  -g, --grouping-strategy
[rmsd|fingerprint|isomorphism|formula|connectivity]
                                  Grouping strategy to use for grouping.
                                  Available options are 'rmsd',
'tanimoto',
                                  'isomorphism', 'formula',
'connectivity'
  -v, --value TEXT                Threshold for grouping strategies.For
rmsd,
                                  it is the rmsd_threshold value.For
Tanimoto,
                                  it is the similarity_threshold
value.For
                                  connectivity, it is the
bond_cutoff_buffer
                                  value.
  -n, --num-grouper-processors INTEGER
                                  Number of processors for grouping.
  --help                          Show this message and exit.
```

Running

```
submit_jobs.py --help
```

gives the following output:

```
Usage: submit_jobs.py [OPTIONS]

  Script for submitting a list of jobs from a .txt file. The .txt file

  contains the names of the files to be submitted. Usage: submit_jobs.py -
f

  filename.txt -c "command" where "command" is the submission command for
all

  files, e.g., "chemsmart sub gaussian -p test -f file sp" where "file" is
the

  placeholder (required) for all filenames in the .txt file.

Options:

  -f, --filename TEXT  .txt file containing the names of files to be

                       submitted.  [required]

  -c, --command TEXT   submission command for all jobs.  [required]

  --help               Show this message and exit.
```

Running

```
wbi_analysis.py --help
```

gives the following output:

```
Usage: wbi_analysis.py [OPTIONS]

    CLI for running wbi analysis using the chemsmart framework.

Options:

  -f, --filename TEXT    Gaussian output file for Wiberg Bond Index
analysis.

                         [required]

  -n, --numbers INTEGER  Atom numbers from which to obtain data from
Wiberg

                         Bond Index output file. 1-indexed.

  --help                 Show this message and exit.
```

Running

```
write_xyz.py --help
```

gives the following output:

```
Usage: write_xyz.py [OPTIONS]

  Script for writing structure to .xyz format. The script can write a
single

  structure to a file or a list of structures, based on 1-indexing. The

  default is to write the last structure in the file.

Options:

  -f, --filename TEXT             Input filename to be converted.

  -i, --index TEXT                 Index of structure to be written to
file.

  -s, --single-file / --no-single-files

                                   To write all structures to a
single .xzy

                                   file, if more than one structure is
present.

                                   Default is to write all structures to a

                                   single file.

  --help                           Show this message and exit.
```